\documentclass[twocolumn,epjc3]{svjour3}  
\smartqed
\usepackage[utf8,latin1]{inputenc}
\newcommand{\qm}[1]{``#1''}
\usepackage{amsmath}
\usepackage{amssymb}
\usepackage{mathabx}
\usepackage{mathrsfs}  
\usepackage{cases}
\usepackage{revsymb}
\usepackage{tabularx}
\usepackage{colortbl}
\usepackage{graphicx}
\usepackage{dcolumn}
\usepackage{xcolor}
\usepackage{bm}
\usepackage{hyperref}
\hypersetup{colorlinks, linkcolor={red},citecolor={blue},urlcolor={blue}}  
\usepackage[square,numbers,sort&compress]{natbib}
\usepackage{hyperref}
\usepackage{stackengine,scalerel}

\newcommand\overstarbf[1]{\ThisStyle{\ensurestackMath{%
  \stackengine{0pt}{\SavedStyle\mathbf{#1}}{\smash{\SavedStyle*}}{O}{c}{F}{T}{S}}}}
\RequirePackage{graphicx}
\journalname{Eur. Phys. J. C}

\begin{document}

\title{Gravitational waves  at the first post-Newtonian order\\ with the Weyssenhoff fluid in  Einstein-Cartan theory}
\titlerunning{Gravitational waves at the first post-Newtonian order with the Weyssenhoff fluid in  Einstein-Cartan theory}      

\author{Emmanuele Battista\thanksref{e1,e2,addr1}
\and
Vittorio De Falco\thanksref{e3,addr2,addr3}}

\thankstext{e1}{e-mail: emmanuele.battista@univie.ac.at}
\thankstext{e2}{e-mail: emmanuelebattista@gmail.com}
\thankstext{e3}{e-mail: vittorio.defalco-ssm@unina.it}

\authorrunning{Battista \& De Falco (2022)}

\institute{Department of Physics, University of Vienna, Boltzmanngasse 5, A-1090 Vienna, Austria \label{addr1}
\and
Scuola Superiore Meridionale, Largo San Marcellino 10, 80138 Napoli, Italy\label{addr2}
\and
Istituto Nazionale di Fisica Nucleare, Sezione di Napoli, Complesso Universitario di Monte S. Angelo, Via Cintia Edificio 6, 80126 Napoli, Italy \label{addr3}}

\date{Received: \today / Accepted: }

\maketitle

\begin{abstract}
The generation of gravitational waves from a  post-Newtonian source endowed with a quantum spin,  modeled by the Weyssenhoff fluid, is investigated in the context of Einstein-Cartan theory at the first post-Newtonian level by resorting to the Blanchet-Damour formalism. After having worked out the basic principles of the hydrodynamics in Einstein-Cartan framework, we study the Weyssenhoff fluid within  the post-Newtonian approximation scheme. The complexity of the underlying dynamical equations suggests to employ a discrete description  via the point-particle limit, a procedure which permits  the  analysis of inspiralling spinning compact binaries. We then provide a first application of our results  by considering  binary neutron star systems. 
\end{abstract}

\section{Introduction}
\label{sec:intro}

Nowadays, gravitational-wave (GW) astronomy represents a fundamental mean to investigate gravity at extreme regimes and offers valuable insights into the  physics of compact objects \cite{Sathyaprakash2009,Bailes2021}, as the best GW candidates are represented by black holes (BHs) and neutron stars (NSs) \cite{LIGOScientific2021psn}. GWs manifest as perturbations of the spacetime and their theoretical description lays solid roots in general relativity (GR). \emph{The GW theory is intimately intertwined with the  two-body problem}. A variety of analytical and numerical techniques has been developed to foretell approximately the dynamics and the corresponding waveforms of compact-object binary systems during their inspiral, plunge, merger, and ringdown stages \cite{Buonanno2014aza,Schmidt2020}. The motion and the radiation of post-Newtonian (i.e., slowly moving, weakly stressed,  and weakly self-gravitating) isolated sources during their early inspiralling stage can be tackled  via the Blanchet-Damour scheme. This framework is built on the pioneering works by Bonnor and collaborators   \cite{Bonnor1959,Bonnor1961,Bonnor1966,Hunter1968} and Thorne \cite{Thorne1980}, and exploits two approximation strategies: the multipolar-post-Minkowskian (MPM) and the post-Newtonian (PN) methods \cite{Blanchet-Damour1986,Blanchet1986}. Other two fundamental GW generation formalisms are the 
Will-Wiseman-Pati approach, which extends the pattern first  developed by Epstein and
Wagoner \cite{Epstein1975}, and the gravitational self-force (GSF) model. The former reckons with the direct integration of the relaxed Einstein equations (DIRE) and  differs from the aforementioned MPM-PN program in the definition of the source multipole moments   \cite{Will1996,Pati2000,Pati2002,Poisson-Will2014}. GSF is based on BH perturbation theory and  explores the dynamics and the radiative phenomena of extreme-mass-ratio inspirals \cite{Barack2018yvs,Berry2019wgg,Pound2019,Pound2021}. The effective-one-body (EOB) framework provides a highly-accurate description of the motion and the gravitational amplitude of coalescing binaries in their late evolution phases \cite{Buonanno1999,Buonanno2000,Damour2011,Damour2016b,Damour2017a,Damour2020a}. A valid support, especially for the demanding task of solving the Einstein field equations in the most extreme regimes, is provided by   numerical relativity (NR). Indeed,  NR simulations are firmly harnessed to predict waveforms through the merger and the ringdown, and validate the approaches used to study binary systems  \cite{Cardoso2014,Healy2019jyf,Boyle2019kee,Ramos-Buades2019uvh,Healy2020vre}. Furthermore, NR is largely exploited together with several phenomenological patterns (Phenom) to perform the fitting and  the parameter estimation of the GW data  \cite{Ajith2008,Ajith2011,Pratten2020,Pratten2020b,Hamilton2021}.

The state of the art on the conservative dynamics consists in the complete set of the equations of motion of point-particle nonspinning binaries at the  fourth post-Newtonian (4PN) order, where the underlying calculations  have been undertaken within three different patterns: the Arnowitt-Deser-Misner (ADM) Hamiltonian formulation of GR \cite{Damour2014j,Damour2016a}, the Fokker-action approach in harmonic coordinates  \cite{Bernard2015,Marchand2017}, and the effective-field-theory model \cite{Foffa2019a,Foffa2019b}; recently, the 5PN and 6PN levels have been worked out modulo a small number of unknowns  \cite{Bini2019a,Bini2020a,Bini2020c,Bini2020r}. In addition, this field is currently investigated via new methodologies making use of tools stemming from effective field theory and modern scattering amplitude programs \cite{Bern2019,Bern2019b,Bern2021dq}. Regarding the GW emission aspects, the 3PN-accurate waveform  has been determined \cite{Blanchet2005b} and a great deal of efforts is being made in the literature to extend  our  present knowledge regarding GW templates of inspiralling compact binaries up to 4PN level  \cite{Marchand2020,Larrouturou2021a,Larrouturou2021b}. In the case of spinning  binary systems, both the radiative aspects and equations of motion have been worked out with  high PN accuracy \cite{Damour2007a,Marsat2012,Bohe2012,Bohe2013c,Marsat2013c,Marsat2014,Levi2014,Bohe2015a,Levi_2016,Cho2021,Cho2022}. 

The huge amount of complementary observational data, triggered by the fast advance of the technological progress, is increasing our capacity to acquire more and more accurate information on gravitational sources and gravity itself. These scientific achievements have motivated us to explore the interplay between quantum and GR effects in GW phenomena via the Einstein-Cartan (EC) theory. Indeed, in a previous paper \cite{Paper1} we have solved,  in the context of EC model,  the GW generation problem at 1PN level by exploiting the Blanchet-Damour formalism, which permits relating  the outgoing radiative gravitational field to the structure and the motion of a spinning PN isolated source. The final solution is encoded by the 1PN-accurate relations for the radiative moments, which   are given in the form of well-defined (compact-support) integral expressions over the stress-energy distribution of the matter field.  This result can be obtained after a detailed analysis regarding the gauge condition, the coordinates covering  the  interior and the exterior zone of the source, and the structure of the torsion tensor, which is supposed to admit a vanishing trace. Furthermore,  the invariance of the  Riemann tensor in the context of the  linearized EC framework must be invoked in order to match the internal and the external fields and write the final expressions of the radiative moments involving the new spin contributions. Hereafter,  \qm{spin} is intended as  the \emph{quantum} (microscopic) angular momentum  of elementary particles \cite{Hehl1976_fundations}. 

In this paper, we apply the abovementioned Blanchet-Damour scheme to a GW source described by the \emph{Weyssenhoff model  of an ideal fluid with spin} \cite{Weyssenhoff1947,Boehmer2006}. This choice is motivated by the fact that it configures as a natural extension of the GR perfect fluid and permits to simplify the demanding calculations framed in EC theory. The hydrodynamical picture of the Weyssenhoff treatment can be obtained by  considering first a perfect fluid, and then by assigning to each \qm{fluid element}, which contains a set of microscopic spin configurations, a value of the spin density tensor via an average procedure \cite{Obukhov1987,Gasperini1998}. The Weyssenhoff fluid has been considered both in cosmological models and astrophysical problems. In the first case, it has been proved that spin interactions in the early stages of the universe  could bring about significant results, such as the avoidance of the big-bang singularity, as well as the reproduction of cosmic inflation and  dark energy mechanisms \cite{Obukhov1987,Gasperini1998,Szydlowski2003,Medina2018}.  In astrophysical settings,  the  spin may avert the spacetime singularity caused by the  gravitational collapse of a star \cite{Ziaie2013,Hashemi2015,Bohmer2018,Luz2019,Hensh2021}. 

In our analysis, the dynamics of the Weyssenhoff fluid is investigated through the PN method, which permits obtaining a hierarchy of PN  mathematical problems to be solved perturbatively order by order starting from the 0PN level. However, this approach still leads to a set of partial and integro-differential dynamical equations, which in general can be  tackled through NR. Therefore, in order to obtain analytical results, we resort to the point-particle procedure, which is a valid pattern widely exploited in the literature \cite{Poisson-Will2014}. In this way, the bodies can be treated as point-like objects and their dynamics is described in terms of ordinary differential equations. This prepares the ground for the analysis of inspiralling spinning compact binaries, which can represent the high-energy astrophysical testbed of our model. 

The article is organized as follows: in Sec. \ref{Sec:1PN-generation-GWs-EC}, we briefly recall the fundamental results of Ref. \cite{Paper1}; in Sec. \ref{Sec:Application-to-Weyssenhoff-fluid}, after having   set  out the main concepts of the hydrodynamics in EC theory, we  deal with  the Weyssenhoff model;  then, we pass from the continuous description to the discrete picture of the fluid by exploiting the point-particle limit, which is worked out in Sec. \ref{sec:Point-particle-limit}; in Sec. \ref{Sec:Application}, we apply our theoretical apparatus   to NS binaries, and provide a first estimate of EC corrections by studying  the ensuing gravitational flux and waveform; eventually, in Sec. \ref{sec:end}, we give a summary of the paper and  draw the conclusions.

\emph{Notations.} We use metric signature  $(-,+,+,+)$. Greek indices take values  $0,1,2,3$, while the Latin ones $1,2,3$. The flat metric is indicated by $\eta^{\alpha \beta }=\eta_{\alpha \beta }={\rm diag}(-1,1,1,1)$. The determinant of the metric $g_{\mu \nu}$ is denoted by $g$. $\epsilon_{kli}$ is   the completely antisymmetric Levi-Civita symbol, whose value is $1$ if $kli$ is an even permutation of $123$. Four-vectors are written as $a^\mu = (a^0,\boldsymbol{a})$ and we employ the following notations: $\boldsymbol{a} \cdot \boldsymbol{b} \equiv \delta_{lk}a^l b^k$, $\vert \boldsymbol{a} \vert \equiv a =  \left(\boldsymbol{a} \cdot \boldsymbol{a}\right)^{1/2}$, and $\left(\boldsymbol{a} \times \boldsymbol{b}\right)^i \equiv \epsilon_{ilk} a^l b^k$. Round (respectively, square) brackets around a pair of indices stands for the usual symmetrization (respectively, antisymmetrization) procedure, i.e., $A_{(ij)}=\frac{1}{2}(A_{ij}+A_{ji})$ (respectively, $A_{[ij]}=\frac{1}{2}(A_{ij}-A_{ji})$).

\section{First post-Newtonian generation of gravitational waves in Einstein-Cartan theory} \label{Sec:1PN-generation-GWs-EC}

We briefly recall the essential information on EC theory and set up the mathematical tools in Sec. \ref{sec:EC_theory}. After that, we introduce the mathematical framework related to the GW generation problem in EC theory at 1PN level  (see Sec. \ref{sec:math_pro}). Then, we outline the procedure, based on the Blanchet-Damour approach, which permits obtaining the approximate solution  (see Sec. \ref{sec:Res_Meth}). Finally, the general expressions of the asymptotic gravitational waveform and the radiated power  are displayed (see Sec. \ref{sec:GW_PR}).

\subsection{Einstein-Cartan theory}
\label{sec:EC_theory}

The EC theory is an extension of GR where both the spin and the mass of matter play
a dynamical role. This model  is defined on a  spacetime $\mathscr{M}$, endowed with a symmetric metric tensor $g_{\alpha \beta}$ and  the most general  metric-compatible affine connection 
\begin{align} 
\Gamma^\lambda_{\mu \nu}=&\hat{\Gamma}^{\lambda}_{\mu \nu}-K_{\mu \nu}^{\phantom{\mu \nu} \lambda},\label{eq:Affine_Connection_1}
\end{align}
where $\hat{\Gamma}^{\lambda}_{\mu \nu}$ denotes the \emph{Christoffel symbols} and $K_{\mu \nu}^{\phantom{\mu \nu} \lambda}= S_{\nu \phantom{\lambda} \mu}^{\phantom{\nu} \lambda}- S^{\lambda}_{\phantom{\lambda} \mu \nu} - S_{\mu \nu}^{\phantom{\mu \nu} \lambda}$ the \emph{contortion tensor}, with $S_{\mu \nu}^{\phantom{\mu \nu} \lambda} \equiv \Gamma^\lambda_{[\mu \nu]} $ dubbed  the \emph{Cartan torsion tensor}. Hereafter, a hat symbol refers to quantities framed in GR.

Given the matter Lagrangian  density,  it is possible to introduce the metric energy-momentum tensor $T^{\alpha \beta}$, \emph{the  spin angular momentum tensor} $\tau_{\gamma}^{\phantom{\gamma}\beta \alpha}$, and the spin energy potential $\mu_{\gamma}^{\phantom{\gamma}\beta \alpha}$  \cite{Hehl1976_fundations,Paper1}. In particular, the tensors $\mu_{\gamma}^{\phantom{\gamma}\beta \alpha}$ and $\tau_{\gamma}^{\phantom{\gamma}\beta \alpha}$ are related in the following way:  
\begin{equation} \label{eq:mu_tensor}
 \mu^{\alpha\beta\gamma}=-\tau^{\alpha\beta\gamma}+\tau^{\beta\gamma\alpha}-\tau^{\gamma\alpha\beta}.
\end{equation}
Another fundamental object is the  total energy-momentum tensor of matter $\mathbb{T}^{\alpha\beta}$,  defined as  \cite{Hehl1976_fundations,Gasperini-DeSabbata}
\begin{equation}
\label{eq:canonical_stress-energy_tensor}
\begin{aligned}
\mathbb{T}^{\alpha\beta}&=T^{\alpha\beta}-\overstarbf{\nabla}_\gamma\left( \mu^{\alpha\beta\gamma}\right),
\end{aligned}
\end{equation}
where $ \overstarbf{\nabla}$  is the \emph{modified covariant derivative} operator, whose action on a generic tensor field of type $(1,1)$ is
\begin{equation}
  \overstarbf{\nabla}_\alpha A^{\mu}_{\phantom{\mu}\nu}=\left(\nabla_\alpha+2S_{\alpha\beta}{}^\beta\right)A^{\mu}_{\phantom{\mu}\nu}. \label{eq:modified-divergence}
\end{equation}
The EC field equations are  ($\chi \equiv 16 \pi G/c^4$)  \cite{Hehl1976_fundations}
\begin{subequations} 
\label{eq:Einstein-Cartan_equations}
\begin{align} 
\hat{G}^{\alpha\beta}&=\frac{\chi}{2}\Theta^{\alpha\beta},\label{eq:hat-G-equals-tilde-T}
\\
\Theta^{\alpha\beta}&\equiv T^{\alpha\beta}+\frac{\chi}{2}\mathcal{S}^{\alpha \beta},\label{eq:tilde-T-alpha-beta}    
\\
\mathcal{S}^{\alpha \beta} & \equiv
    -4\tau^{\alpha\gamma}{}_{[\delta}\tau^{\beta\delta}{}_{\gamma]}-2\tau^{\alpha\gamma\delta}\tau^\beta{}_{\gamma\delta}+\tau^{\gamma\delta\alpha}\tau_{\gamma\delta}{}^\beta
    \nonumber \\
    &+\frac{1}{2}g^{\alpha\beta}\left(
    4\tau_{\mu}^{\phantom{\mu}\gamma}{}_{\phantom{\gamma}[\delta}\;\tau^{\mu\delta}{}_{\gamma]}+\tau^{\mu\gamma\delta}\tau_{\mu\gamma\delta}\right),
\label{Eq:S_alphabeta}
\end{align}
\end{subequations}
where the combined energy-momentum tensor $\Theta^{\alpha\beta}$ satisfies 
\begin{equation} \label{eq:nabla_tilde_T}
    \hat{\nabla}_\beta \Theta^{\alpha\beta}=0.
\end{equation}
The matter source's dynamical equations   can be obtained via the generalized conservation laws of energy-momentum and  angular momentum, which read as, respectively,\footnote{It is possible to show that Eqs. (\ref{eq:nabla_tilde_T}) and (\ref{eq:generalized-conservation-laws-energy-ang-momentum}) are not independent (see Ref. \cite{Hehl1976_fundations} for further details).} \cite{Hehl1976_fundations,Gasperini-DeSabbata}
\begin{subequations}
\label{eq:generalized-conservation-laws-energy-ang-momentum}
\begin{align} 
\overstarbf{\nabla}_\nu  \mathbb{T}_{\mu}^{\phantom{\mu}\nu} =& 2 \mathbb{T}_{\lambda}^{\phantom{\lambda}\nu} S_{\mu \nu}^{\phantom{\mu \nu }\lambda} -  \tau_{\nu \rho}^{\phantom{\nu \rho}\sigma} R_{\mu \sigma}^{\phantom{\mu \sigma}\nu \rho},
\label{eq:conservation-law-energy-momentum}\\
\overstarbf{\nabla}_\lambda \tau_{\mu \nu}^{\phantom{\mu \nu}\lambda} =& \mathbb{T}_{[\mu \nu]},
\label{eq:conservation-law-angular-momentum}
\end{align}
\end{subequations}
the Riemann tensor  being given by
\begin{align} \label{eq:EC-Riemann-tensor}
R^{\mu}_{\phantom{\mu}\nu \rho \sigma}&= \partial_\rho \Gamma^{\mu}_{\sigma\nu }-\partial_\sigma \Gamma^{\mu}_{\rho \nu} + \Gamma^{\mu}_{\rho \alpha} \Gamma^{\alpha}_{ \sigma\nu}-\Gamma^{\mu}_{\sigma\alpha} \Gamma^{\alpha}_{\rho\nu}
\nonumber \\
&=   \hat{R}^{\mu}_{\phantom{\mu}\nu \rho \sigma} + \hat{\nabla}_\sigma K_{\rho \nu}^{\phantom{\rho \nu}\mu}-\hat{\nabla}_\rho K_{\sigma \nu}^{\phantom{\sigma \nu}\mu}
 \nonumber \\
 &+ K_{\rho \alpha}^{\phantom{\rho \alpha}\mu}K_{\sigma \nu}^{\phantom{\sigma \nu}\alpha}- K_{\sigma \alpha}^{\phantom{\sigma \alpha}\mu}K_{\rho \nu}^{\phantom{\rho \nu}\alpha}.
\end{align}

\subsection{The mathematical problem}
\label{sec:math_pro}

The GW generation in EC theory is described by the following well-posed mathematical problem \cite{Paper1}:
\begin{subnumcases}{\label{eq:math-problem-EC2}}
\Box \mathfrak{h}^{\mu\nu}=\chi \tilde{\mathfrak{T}}^{\mu \nu}, & \label{eq:proa-EC2}\\ 
\partial_\lambda \mathfrak{h}^{\alpha \lambda}=0, & \label{eq:prob-EC2}\\ 
\lim \limits_{|\boldsymbol{x}|\to +\infty}\mathfrak{h}^{\alpha\beta}(t,\boldsymbol{x})=0, \qquad \mbox{for}\ t\le -\mathcal{T}, & \label{eq:proc-EC2}\\ 
\partial_t\mathfrak{h}^{\alpha\beta}(t,\boldsymbol{x}) =0,\hspace{1.55cm} \mbox{for}\ t\le -\mathcal{T}, & \label{eq:prod-EC2} 
 \end{subnumcases}
where $\mathfrak{h}^{\alpha\beta} \equiv \sqrt{-g}g^{\alpha \beta}-\eta^{\alpha\beta}$, $\Box \equiv \eta^{\alpha \beta} \partial_\alpha \partial_\beta$,  and 
\begin{equation} \label{eq:tilde-tau-alpha-beta-def-1}
 \tilde{\mathfrak{T}}^{\alpha\beta}\equiv \left(-g\right)\, \Theta^{\alpha\beta}+\frac{1}{\chi}\Lambda^{\alpha\beta},
 \end{equation}
 is the \emph{effective stress-energy pseudotensor}  encompassing  both the matter fields, described by  $\Theta^{\alpha\beta}$, and the  effective gravitational source  term $\Lambda^{\alpha\beta}=\Lambda^{\alpha\beta}(\mathfrak{h})$, which includes all the nonlinearities of EC field equations.  The harmonic gauge  (\ref{eq:prob-EC2}) can be imposed provided we further require  \cite{Paper1}
\begin{equation} \label{eq:simplification-of-harmonic-gauge}
    S^{\alpha \mu}_{\phantom{ \alpha \mu}  \mu}=0.
\end{equation}

The gravitational source is supposed to be confined to the region $\Omega$, defined by 
\begin{equation}\label{eq:Omega-region}
\Omega=\left\{\boldsymbol{x}\in\mathbb{R}^3:|\boldsymbol{x}|\leq \bar{d}\right\},   
\end{equation}
with $\bar{d}$  the typical size of the source. This condition ensures, on the one hand, that $T^{ \alpha \beta}$, $\tau_{\gamma}^{\phantom{\gamma}\beta \alpha}$, and $\Theta^{\alpha\beta}$ are smooth functions in $\mathbb{R}^4$ having a spatially compact support in $\Omega$,  and, on the other, that  the torsion tensor vanishes outside $\Omega$.

We deal with \emph{PN sources in EC theory}, for which the spatial domain $\mathbb{R}^3$ can be decomposed as $\mathbb{R}^3=\mathfrak{D}_e\cup\mathfrak{D}_i$, where the set $\mathfrak{D}_i$ is the \emph{near zone} and covers entirely the source, while $\mathfrak{D}_e$ encompasses the external weak-field  region of the source and  is called \emph{exterior zone}. For a PN source these two domains intersect in the  \emph{overlapping  region} $\mathfrak{D}_o$.
Finally, the spatial region where a detector apparatus is located is known as \emph{wave zone}.

\subsection{Resolution method}
\label{sec:Res_Meth}

The (approximate) solution of problem (\ref{eq:math-problem-EC2}) can be worked out as follows \cite{Paper1,Blanchet-Damour1989}. First of all, two preliminary steps need to be carried out in the overlapping region $\mathfrak{D}_o$: (1) calculating the multipole (re)expansion of the 1PN series of the inner metric, which can be expressed in terms of the \emph{source multipole moments} $I_L$ and $J_L$\footnote{We have used the \emph{multi-index notation}, where $L$ denotes the multi-index $i_1i_2\dots i_l$, hence $I_L=I_{i_1i_2\dots i_l}$ \cite{Blanchet-Damour1986}.}; (2) evaluating the 1PN (re)expansion of the MPM external metric, which depends on the \emph{canonical multipole moments} $M_L$ and $S_L$.  Eventually, $I_L,J_L$ and  $M_L,S_L$  are related to the \emph{radiative moments} $U_L, V_L$ (representing the physical observables in the wave zone) by exploiting the matching procedure. At 1PN order, the relation between $I_L, J_L$ and  $U_L, V_L$ is given by (with $l \geq 2$) \cite{Paper1}
\begin{subequations}
\label{eq:radiative-source_1PN}
\begin{align} 
       U_L(u) &= \overset{(l)}{I}_L(u)+ {\rm O}(c^{-3}),
 \\
       V_L(u) &= \overset{(l)}{J}_L(u) + {\rm O}(c^{-2}),
\end{align}
\end{subequations}
the superscript $(l)$ denoting the $l$-th time derivative with respect to the variable $u$, and
\begin{subequations}
\label{eq:I_L and J_L-torsion}
\begin{align} 
I_L(u) & = \int {\rm d}^3 \boldsymbol{y} \, y_{\langle L\rangle } \sigma(\boldsymbol{y},u) 
\nonumber\\
& + \dfrac{1}{2(2l+3)}\dfrac{1}{c^2} \dfrac{{\rm d}^2}{{\rm d}u^2} \int {\rm d}^3 \boldsymbol{y} \, y_{\langle L\rangle } \,  \boldsymbol{y}^2 \sigma(\boldsymbol{y},u) 
\nonumber \\
& - \dfrac{4(2l+1)}{(l+1)(2l+3)}\dfrac{1}{c^2} \dfrac{{\rm d}}{{\rm d}u} \int {\rm d}^3 \boldsymbol{y} \, y_{\langle iL\rangle } \sigma_i (\boldsymbol{y},u)
\nonumber \\
& + {\rm O}(c^{-4}), \qquad \qquad   (l \geq 0),
 \label{eq:I_L_torsion}
 \\
 J_L(u) & = \int {\rm d}^3 \boldsymbol{y} \, \epsilon_{ab\langle i_l} \widecheck{y}_{L-1\rangle a} \sigma_b(\boldsymbol{y},u) 
 \nonumber \\
 & +  {\rm O}(c^{-2}), \qquad \qquad   (l \geq 1),
 \label{eq:J_L_torsion}
\end{align}
\end{subequations}
where
\begin{equation}
\label{eq:def_sigma-sigma_i}
\sigma \equiv \dfrac{\Theta^{00}+\Theta^{kk}}{c^2}, \qquad \sigma_i \equiv \dfrac{\Theta^{0i}}{c},
\end{equation}
and $y_{\langle L\rangle } = \widecheck{y}_{L}$ stands for the symmetric-trace-free (STF) projection of $y_L$. 

\subsection{Asymptotic gravitational waveform and radiated power}
\label{sec:GW_PR}

 Given a set $X^\mu=(cT,\boldsymbol{X})$ of radiative coordinates, the external metric can be put in the so-called \emph{radiative form}, where its coefficients admit an asymptotic expansion in powers of $\mathcal{R}^{-1}$ at future null infinity (i.e., $\mathcal{R} \equiv \vert \boldsymbol{X} \vert \to \infty$ with $\mathcal{U} \equiv T-\mathcal{R}/c$ and $\boldsymbol{\mathcal{N}} \equiv \boldsymbol{X}/\mathcal{R}$ fixed) \cite{Blanchet1986,Blanchet-Damour1989,Blanchet1995,Blanchet2014,Maggiore:GWs_Vol1}. The \emph{asymptotic waveform} $\mathscr{H}^{\rm TT}_{ij}(X)$ describing the outgoing radiation is defined as the transverse-traceless (TT) projection of the leading $\mathcal{R}^{-1}$ term of such an expansion. At 1PN order, it reads as 
\begin{align} \label{eq:gravitational_wave_amplitude}
\mathscr{H}_{ij}^{\rm TT}(X^\mu) & = \dfrac{2G}{c^4 \mathcal{R}} \mathscr{P}_{ijkl}(\boldsymbol{\mathcal{N}})  \Biggr\{ U_{kl}(\mathcal{U}) 
\nonumber \\
& + \dfrac{1}{c} \left[  \dfrac{1}{3} \mathcal{N}_a U_{kla}(\mathcal{U}) +\dfrac{4}{3} \epsilon_{ab(k} V_{l)a}(\mathcal{U}) \mathcal{N}_b \right]
\nonumber \\
& +\dfrac{1}{c^2} \left[\dfrac{1}{12}\mathcal{N}_{ab} U_{klab}(\mathcal{U})
+ \dfrac{1}{2} \epsilon_{ab(k} V_{l)ac}(\mathcal{U}) \mathcal{N}_{bc} \right]
\nonumber \\
& + {\rm O}(c^{-3}) \Biggr\},
\end{align} 
where  
\begin{align}
    \mathscr{P}_{ijkl} (\boldsymbol{\mathcal{N}}) & \equiv \mathscr{P}_{ik}\mathscr{P}_{jl}-\dfrac{1}{2}\mathscr{P}_{ij}\mathscr{P}_{kl},
    \nonumber \\
    \mathscr{P}_{ij} (\boldsymbol{\mathcal{N}}) & \equiv \delta_{ij}-\mathcal{N}_i \mathcal{N}_j,
\end{align}
 $\mathscr{P}_{ijkl}(\boldsymbol{\mathcal{N}})$ being the TT projection operator onto the plane orthogonal to $\boldsymbol{\mathcal{N}}$.

Starting from the results contained in Refs. \cite{Maggiore:GWs_Vol1,Poisson-Will2014,MTW}, we have proved that the standard GR formula of  the  \emph{total radiated power} $\mathcal{F}$ (also called,  in the astrophysics literature,  \emph{total gravitational luminosity or flux of the source}) is valid also in EC theory.  Therefore, at 1PN order,  the total energy radiated per unit time, expressed as a function of the retarded time $\mathcal{U}$, reads as
\begin{align} \label{eq:power-radiated-1PN}
    \mathcal{F}(\mathcal{U})  &= \dfrac{G}{c^5} \Biggl\{\dfrac{1}{5}\overset{(1)}{U}_{ij}(\mathcal{U})\overset{(1)}{U}_{ij}(\mathcal{U}) 
   \nonumber \\
 &+ \dfrac{1}{c^2} \Biggl[ \dfrac{1}{189} \overset{(1)}{U}_{ijk}(\mathcal{U}) \overset{(1)}{U}_{ijk}(\mathcal{U})
 \nonumber \\
&+ \dfrac{16}{45} \overset{(1)}{V}_{ij}(\mathcal{U})\overset{(1)}{V}_{ij}(\mathcal{U}) \Biggr]
+ {\rm O}(c^{-4}) \Biggr \}.
\end{align}

\section{The semiclassical spin fluid and its post-Newtonian approximation}
\label{Sec:Application-to-Weyssenhoff-fluid}

Having set out the main aspects of the Blanchet-Damour approach in EC theory, we consider, as a first approach to the description  of  spin effects inside matter,  the class of semiclassical spin fluid models. Before getting to the heart of the discussion, in Sec. \ref{sec:rel_hyd} we first present the main  principles of the hydrodynamics in EC theory by making use of the hypothesis \eqref{eq:simplification-of-harmonic-gauge}. In Sec. \ref{sec:Weyssenhof}, we consider the Weyssenhoff fluid, which represents  one of the most common frameworks studied  in the literature. Finally, in Sec. \ref{Sec:PN-Hydrodynamics}, we investigate the Weyssenhoff fluid dynamics within the PN approximation scheme. 

\subsection{Hydrodynamics in Einstein-Cartan theory under the hypothesis $S_{\alpha \beta}{}^\beta=0$}
\label{sec:rel_hyd}

In this section, we introduce the basic concepts underlying the  hydrodynamics in EC theory. 

The analysis of a fluid in EC framework turns out to be more complex than in the GR case, but   the assumption (\ref{eq:simplification-of-harmonic-gauge}) entails a great simplification. Indeed, the modified covariant derivative (\ref{eq:modified-divergence}) becomes
\begin{equation} \label{eq:nabla_star&nabla}
    \overstarbf{\nabla}_\lambda \left( \cdot \right)=\nabla_\lambda \left( \cdot \right)=\hat{\nabla}_\lambda \left( \cdot \right)-K_{\lambda \phantom{\cdot}\cdot}^{\phantom{\cdot}\cdot}\left( \cdot \right),
\end{equation}
 and the EC covariant divergence of a generic vector $A^\mu$ assumes the same form as in GR, namely
\begin{equation}\label{eq:nabla-mu-u-mu}
\nabla_\mu A^\mu = \hat{\nabla}_\mu A^\mu -K_{\mu\lambda}{}^\mu A^\lambda= \hat{\nabla}_\mu A^\mu. 
\end{equation}

The spacetime position of each fluid element  is labelled in a system of coordinates by $x^\mu=(ct,\boldsymbol{x})$. The description of the kinematic properties of the fluid can be performed in terms of two fundamental quantities: the  timelike
four-velocity $u^\mu=\frac{{\rm d}x^\mu}{{\rm d}\lambda}$ (where $\lambda$ denotes the proper time of an observer comoving with the fluid) and the four-acceleration $a^\mu=u^\nu\nabla_\nu u^\mu$.

It is useful to introduce the projection operator on the spatial hypersurface orthogonal to the timelike  four-velocity $u^\mu$ 
\begin{equation} \label{eq:projection-operator}
\mathcal{P}^{\mu\nu}= \frac{u^\mu u^\nu}{c^2}+g^{\mu\nu},    
\end{equation}
and the \emph{EC substantial derivative} of the vector $\xi^\mu$ along  $u^\mu$    \cite{Weyssenhoff1947}
\begin{align} \label{eq:MATDER}
\dot{\xi}^\mu\equiv u^\nu \nabla_\nu \xi^\mu.
\end{align}
The latter is the natural extension of the concept used in GR and in classical continuous mechanics, which allows to define the \emph{EC time derivative for densities} \cite{Weyssenhoff1947}
\begin{align} \label{eq:MATDERDEN}
\mathscr{D}\xi^\mu  &\equiv \nabla_\nu \left(u^\nu \xi^\mu\right)=\dot{\xi}^\mu+ \xi^\mu\hat{\nabla}_\nu u^\nu.
\end{align}
The above derivative operator is applied to objects which are densities (e.g., the spin density tensor, which will be introduced in Sec. \ref{sec:Weyssenhof}) and has a precise physical meaning. Indeed, its definition relies on the fact  that the densities must be not only transported along the fluid worldlines, but it must be also taken into account how the volumes transform during the dynamical evolution\footnote{By defining the \emph{expansion tensor} $\vartheta_{\alpha\beta} \equiv \mathcal{P}_\alpha^\mu\mathcal{P}_\beta^\nu \nabla_{(\nu} u_{\mu)}$, it is easy to prove that  the \emph{expansion scalar} $\vartheta \equiv \vartheta^\alpha_{\ \alpha}$ assumes the same form as in GR \cite{Rezzolla2013}, i.e., $\vartheta=2\hat{\nabla}_\alpha u^\alpha$. This quantity  describes how the fluid volume changes during the motion.} (see Ref. \cite{Weyssenhoff1947}, for more details). 

The metric stress-energy tensor modeling the relativistic perfect fluid which includes also spin contributions, can be generally written as  
\begin{align}\label{eq:metric-stress-energy-tensor-general}
T^{\alpha\beta} &=T^{\alpha\beta}_{\rm perf}+ \Phi^{\alpha\beta}, 
\nonumber \\
T^{\alpha\beta}_{\rm perf} &=e\frac{u^\mu u^\nu}{c^2} + \mathcal{P}^{\mu\nu}P,
\end{align}
where  $\Phi^{\alpha\beta}$ \emph{is a symmetric tensor containing torsion terms  and whose explicit form depends on the chosen spin model}, $P$  denotes the isotropic fluid pressure, and 
\begin{align} \label{eq:hydrodynamics-total-en-dens-general-express}
e &= \rho c^2+\varepsilon,
\end{align}
 the total energy density, which, in turn, depends on   the rest-mass density $\rho$ and   the internal energy density $\varepsilon$. We note that  $e$ can be functionally split in the sum of $\rho c^2$ and $\varepsilon$ due to the perfect fluid hypothesis. 
 
Upon introducing the rest-mass density current
\begin{equation}
    J^\mu=\rho u^\mu,
\end{equation}
we can write the conservation equation for the rest mass as (cf. Eq. (\ref{eq:nabla-mu-u-mu}))
\begin{equation} \label{eq:cons-energy}
\hat{\nabla}_\mu J^\mu=0, \qquad \Leftrightarrow \qquad
\mathscr{D}\rho=0.
\end{equation} 
Furthermore, Eq.  (\ref{eq:nabla_star&nabla}) permits simplifying  the expressions of the canonical stress-energy tensor (\ref{eq:canonical_stress-energy_tensor}) and of the combined stress-energy tensor (\ref{eq:tilde-T-alpha-beta}), which read as, respectively,
\begin{subequations} \label{eq:canonical&combined-stress-energy-tensor}
\begin{align}
\mathbb{T}^{\alpha\beta} =& T^{\alpha\beta}_{\rm perf}+ \Phi^{\alpha\beta}-\nabla_\lambda(\mu^{\alpha\beta\lambda}),
\\
\Theta^{\alpha\beta} =& T^{\alpha\beta}_{\rm perf}+ \Phi^{\alpha\beta}+\frac{\chi}{2}\mathcal{S}^{\alpha\beta}.
\end{align}
\end{subequations}
Therefore, from Eq.  (\ref{eq:conservation-law-energy-momentum}), we have 
\begin{align} \label{eq:conservation-law-energy-momentum_2}
&\nabla_\nu \left(T^{\mu\nu}_{\rm perf}+ \Phi^{\mu\nu}-\nabla_\lambda \mu^{\mu\nu\lambda}\right) =2T^{\lambda\nu}_{{\rm perf}}\, S^{\mu }_{\phantom{ \mu} \nu\lambda}
\nonumber\\
&+ 2\Phi_{\lambda\nu}S^{\mu\nu\lambda} -2\nabla_\gamma(\mu_{\lambda\nu}{}^\gamma)S^{\mu\nu\lambda}-\tau_{\nu\rho\sigma}R^{\mu\sigma\nu\rho}.
\end{align}

The dynamics of the perfect fluid can be examined via Eqs.   \eqref{eq:cons-energy} and \eqref{eq:conservation-law-energy-momentum_2}, supplemented by Eq. (\ref{eq:conservation-law-angular-momentum}). In this way, we obtain a system of highly-non-linear  differential equations, dubbed   \emph{cardinal equations of the hydrodynamics in EC theory}, which can be written as 
\begin{subnumcases}{\label{system of general matter equations}}
\hat{\nabla}_\mu J^\mu=0, \label{eq:CONSEN}\\
\nabla_\nu T^{\mu\nu}_{\rm perf}+\nabla_\nu \Phi^{\mu\nu}-\nabla_\nu\nabla_\lambda(\mu^{\mu\nu\lambda})\notag\\
 \quad-2T^{\lambda\nu}_{{\rm perf}}\, S^{\mu }_{\phantom{ \mu} \nu\lambda}-2\Phi_{\lambda\nu}S^{\mu\nu\lambda}\notag\\
\quad+2\nabla_\gamma(\mu_{\lambda\nu}{}^\gamma)S^{\mu\nu\lambda}+\tau_{\nu\rho\sigma}R^{\mu\sigma\nu\rho} =0,
\label{eq:EUL}\\
\nabla_\lambda \tau_{\mu \nu}^{\phantom{\mu \nu}\lambda} -\mathbb{T}_{[\mu \nu]}=0. \label{eq:TENS} 
\end{subnumcases}

Equation \eqref{eq:EUL} can be evaluated  in the directions orthogonal (by means of the projection operator (\ref{eq:projection-operator})) and parallel to   the fluid four-velocity. The former components give the \emph{Euler equation in EC theory}, while the latter the \emph{energy-balance law}. Furthermore, Eq. (\ref{eq:TENS}) leads to the  \emph{rotational equations} of  the fluid motion. 

The system  \eqref{system of general matter equations}  comprises, in general, 11 independent equations and  at most 26 unknowns, which are represented by: the 3 components of the fluid four-velocity (the fourth one is constrained by the normalization condition $u^\mu u_\mu =-c^2$), the rest-mass density $\rho$, the internal energy density $\varepsilon$, the pressure $P$, and  (at most) the 20 independent components of the spin angular momentum tensor $\tau_\lambda^{\ \mu\nu}$. Indeed,   the independent components of the torsion tensor $S_\lambda^{\ \mu\nu}$    have been lowered to 20 by virtue of the hypothesis  \eqref{eq:simplification-of-harmonic-gauge} \cite{Paper1} and hence also $\tau_\lambda^{\ \mu\nu}$ has (at most) 20 independent components. 
This is in agreement with the physical content of the EC theory, 
where the torsion tensor is the geometrical counterpart of the spin of matter, encoded by the tensor $\tau_\lambda^{\ \mu\nu}$. In the worst case, $S_\lambda^{\ \mu\nu}$ can be determined once  the 20 components of  $\tau_\lambda^{\ \mu\nu}$ are known, but, as we will see in the next section, this number  can  be drastically reduced by assigning a suitable functional form for $\tau_\lambda^{\ \mu\nu}$.

It is clear that the system \eqref{system of general matter equations} is in general not \emph{closed}, since there is no  balance between the number of independent equations and  unknowns. However, this not  a drawback of the model, which, on the contrary, can reveal a rich dynamical structure. In order to supply the missing equations and  correctly characterize the structure of the fluid under investigation, \emph{constitutive equations} must be added to the system \eqref{system of general matter equations}, e.g.,   $P=P\left(\rho,\varepsilon\right)$, $P=P(\varepsilon)$, $\rho=\rho(\varepsilon)$. 

A fundamental and general aspect of our approach relies on the possibility of describing the spin effects inside the fluid by exploiting all kinds of (relativistic) spin models. Indeed, the heart of this pattern consists in providing the functional form of $\Phi^{\alpha\beta}$ and $\tau^{\alpha\beta\gamma}$, which is equivalent to assign $T^{\alpha \beta}$ and $\mu^{\alpha\beta\gamma}$. In general, these two unknowns play a crucial physical and geometrical role in the EC gravity framework. Indeed, we can distinguish the following two classes:
(1) \emph{spin-geometry}, represented by either $\tau^{\alpha\beta\gamma}$ or $\mu^{\alpha\beta\gamma}$ (skew-symmetric quantities), which rule the torsion field $S_{\mu\nu}{}^\lambda$; (2) \emph{mass-geometry}, described by either $\Phi^{\alpha\beta}$ or $T^{\alpha \beta}$ (symmetric quantities), which shape the metric tensor $g_{\mu\nu}$.

\subsection{The Weyssenhoff fluid}
\label{sec:Weyssenhof}

This section is devoted to the description of the Weyssenhoff fluid. In Sec. \ref{sec:hystory}, we retrace the historical ideas behind Weyssenhoff approach, since they are useful for its  full comprehension. In Sec. \ref{sec:model}, we present the model and the related dynamical equations. We will see that it can be readily analyzed within our general framework put forth in Sec. \ref{sec:rel_hyd}. A short digression on the first thermodynamic law is contained in Sec. \ref{Sec:first-thermodyanamic-law}.

\subsubsection{Historical introduction to the fluid with spin}
\label{sec:hystory}

The study of matter in its microphysical aspects can be tackled  following different strategies framed either  in quantum models or from the standpoint of classical vision. Due to the arguments developed in this paper, it is more suitable to follow the latter approach, where the structure of physical systems is derived starting from relativistic theories. 

We consider what in the early literature was dubbed  \emph{free spin-particle}, namely a material particle endowed with spin and on which no force, apart from the gravitational pull, acts \cite{Mathisson1937a,Mathisson1937b,Weyssenhoff1947}. We pursue the route paved by the \qm{Krakow school} in the years 1937--1947, characterized by the prolific scientific activity of several Polish physicists, like M. Mathisson, J. Luba{\'n}ski, J. Weyssenhoff, and A. Raabe. In 1927, Einstein and Gromer derived the dynamical equations of a free particle from the equations of the gravitational field as the singularities in this field \cite{Einstein1927}. In 1937, inspired by these ideas, Mathisson and Luba{\'n}ski deduced the equations of motion of a free spin-particle through a variational principle  in  the case of a   linearized gravitational field \cite{Mathisson1937a,Mathisson1937b,Lubanski1937,Sredniawa1980}. This new dynamics exhibits two peculiarities: (1) in the classical limit, it does not reduce to the Newton laws of motion, since it keeps an additional term depending on the internal angular momentum or spin of the particle; (2) the presence of spin generates dynamical equations of the third order.  However, Mathisson did not immediately realize that his results were equivalent to those previously published in 1926 by the Russian physicist Frenkel \cite{Frenkel1926}. Indeed, the latter author, after a discussion with Pauli, who showed him a letter by Thomas on the famous \qm{precession factor $1/2$} \cite{Thomas1927}, became interested in deriving the equations of motion for a spinning electron, which coincide exactly with those of a free spin-particle only if the terms depending on the electromagnetic field vanish. In 1947, after the Second World War,   Weyssenhoff and Raabe proposed a third different method to obtain the equations of a free spin-particle. These authors showed that their results agree with those derived by Frenkel and Mathisson \cite{Weyssenhoff1947}. However, their approach is more rigorous, since it is constructed by exploiting fundamental concepts from continuous mechanics and GR. In addition, the final equations are presented in a much simpler form thanks to the introduction of the linear energy-momentum four-vector \cite{Weyssenhoff1947}. 

The last approach is the most   used in the literature to derive the equations of the incoherent spinning fluid. For this reason, it is also called \emph{Weyssenhoff(-Raabe) fluid}. Having defined $s^{\alpha\beta}$ as the spin density per unit rest-volume, we split it into two three-dimensional vectors \begin{subequations}
\begin{align}
\boldsymbol{s}&=(s^{23},s^{31},s^{12}),\\
\boldsymbol{q}&=(s^{10},s^{20},s^{30}).
\end{align}
\end{subequations}
The fundamental hypothesis behind the Weyssenhoff fluid is that the vector $\boldsymbol{q}$ vanishes in the rest system of the fluid, which translates in having the covariant relation \cite{Frenkel1926,Weyssenhoff1947}
\begin{equation} \label{eq:FC}
    s^{\alpha\beta}u_\beta=0,
\end{equation}
since the fluid velocity $u^\alpha=(u^0,\boldsymbol{u})$ has vanishing spatial components when evaluated in the fluid rest frame. This hypothesis, also known in the literature as \emph{Frenkel condition} since it was first employed by Frenkel, arises from the following practical needs:  (1) simplifying and closing the set of differential equations underlying the dynamics by matching the numbers of unknowns and equations; (2)  letting $\boldsymbol{s}$ be the vector which  encodes the real physical degrees of freedom, since $\boldsymbol{q}$ can be gauged away by an appropriate choice of a coordinate system.  

Equation \eqref{eq:FC} can be written in three-dimensional form  as
\begin{align}
&\boldsymbol{q}=\frac{1}{c}\boldsymbol{u}\times \boldsymbol{s}, 
\end{align}
from which it follows 
\begin{align}
    \boldsymbol{q}\cdot\boldsymbol{u}=0.
\end{align}
The vectors $\boldsymbol{s}$ and $\boldsymbol{q}$ have the same transformation properties of the magnetic and the electric fields, respectively. Therefore,  the Frenkel condition physically tells us that $s^{\alpha\beta}$ is \qm{purely magnetic}, meaning that its electric component vanishes in the rest-frame coordinate system.

\subsubsection{The model}
\label{sec:model}

We consider the  Weyssenhoff semiclassical model of  a   neutral spinning perfect fluid  
in the framework of EC theory \cite{Hehl1976_fundations,Ray1982a,Ray1982b,Griffiths1982,deRitis1983,Obukhov1987,Boehmer2006}.

Following the general approach devised in Sec. \ref{sec:rel_hyd}, a specific fluid can be characterized by assigning the spin angular momentum tensor $ \tau_{\alpha\beta}{}^\gamma$ and the symmetric tensor $\Phi^{\alpha \beta}$. The former, is given by \cite{Obukhov1987}
\begin{align}
\tau_{\alpha\beta}{}^\gamma&=s_{\alpha\beta}u^\gamma,
\label{eq:spin-tensor-fluid}
\end{align}
 $s_{\alpha\beta}=s_{[\alpha\beta]}$ being the spin density tensor, which  is constrained, due to the hypothesis (\ref{eq:simplification-of-harmonic-gauge}), to  satisfy 
\begin{equation} \label{eq:Frenkel_condition}
  \tau_{\alpha\beta}{}^\beta= s_{\alpha\beta}\,u^\beta=0,  
\end{equation}
i.e., the \emph{Frenkel condition} (cf. Eq. (\ref{eq:FC})). This assumption permits several considerable simplifications. First of all, the torsion and contortion tensors can be written as
\begin{subequations}
\label{eq:torsion&contorsion&tau_simplified}
\begin{align} 
    S_{\mu \nu}^{\phantom{\mu \nu}\lambda} =& \dfrac{\chi}{2} \left(\tau_{\mu \nu}^{\phantom{\mu \nu}\lambda} \right),\label{eq:torsion&tau-simplified} 
    \\
      K_{\mu \nu}^{\phantom{\mu \nu}\alpha}=&\dfrac{\chi}{2} \left(-\tau_{\mu \nu}^{\phantom{\mu \nu}\alpha} +\tau_{\nu \phantom{\alpha}\mu}^{\phantom{\nu}\alpha}-\tau^\alpha_{\phantom{\alpha}\mu \nu} \right), 
    \label{eq:contorsion&tau-simplified}  
\end{align}
\end{subequations}
respectively; furthermore,  the EC time derivative of $s_{\mu \nu}$ does not involve contortion terms  and reads as (cf.  Eq.  (\ref{eq:MATDERDEN}))
\begin{align}\label{eq:substantial_derivative_1}
    \mathscr{D}s_{\mu \nu} = \hat{\nabla}_\lambda \left( s_{\mu \nu} u^\lambda \right);
\end{align}
finally, we can write the useful identity 
\begin{equation}
u^\mu \left( \mathscr{D}s_{\mu \nu}\right)= -a^\mu s_{\mu \nu},
\end{equation}
where the acceleration $a^\mu$ is the same as in GR, i.e., 
\begin{equation} \label{eq:acceleration-fluid-simplified}
 a^\mu =  u^\lambda \hat{\nabla}_\lambda u^\mu =\dot{u}^\mu. 
\end{equation}
The tensor $\Phi^{\alpha\beta}$ of the Weyssenhoff fluid is given by \cite{Obukhov1987}
\begin{align}
\Phi^{\alpha\beta} &= 2 \left(\dfrac{u_\mu u^\gamma}{c^2}-\delta^\gamma_\mu\right)\hat{\nabla}_\gamma\left[s^{\mu(\alpha}u^{\beta)}\right]
\nonumber \\
&- \chi \left(s^2 u^\alpha u^\beta + c^2 s^{\alpha}_{\phantom{\alpha}\lambda} s^{\beta \lambda}\right), 
\label{eq:Phi_fluid}
\end{align}
where $s^2 \equiv s^{\alpha \beta}s_{\alpha \beta}$ is the  spin density scalar, satisfying the condition $\mathscr{D}s^2=0$. Therefore, from Eq. \eqref{eq:Phi_fluid}, jointly with Eqs. \eqref{Eq:S_alphabeta}, \eqref{eq:metric-stress-energy-tensor-general}, and \eqref{eq:spin-tensor-fluid},  we have
\begin{align} 
        T^{\alpha \beta} & = e \dfrac{u^\alpha u^\beta}{c^2}+ \mathcal{P}^{\alpha\beta} P 
\nonumber \\
         & + 2 \left(\dfrac{u_\mu u^\gamma}{c^2}-\delta^\gamma_\mu\right)\hat{\nabla}_\gamma\left[s^{\mu(\alpha}u^{\beta)}\right] 
\nonumber  \\
         & - \chi \left(s^2 u^\alpha u^\beta + c^2 s^{\alpha}_{\phantom{\alpha}\lambda} s^{\beta \lambda}\right), \label{eq:T_alpha_beta_fluid} 
\\
\mathcal{S}^{\alpha \beta} &=2c^2 s^{\alpha}_{\phantom{\alpha}\lambda} s^{\beta \lambda} +s^2  u^\alpha u^\beta -\dfrac{1}{2}s^2 c^2 g^{\alpha \beta}, \label{eq:S-tensor-fluid}
\end{align}
and hence the explicit expressions  of the canonical and the combined energy-momentum tensors are, respectively, 
(cf.  Eq.  \eqref{eq:canonical&combined-stress-energy-tensor}) 
\begin{subequations}
\begin{align} 
\mathbb{T}^{\alpha \beta}&= \left( e u^\alpha - 2 a_\sigma s^{\sigma \alpha} \right)\dfrac{u^\beta}{c^2}+\mathcal{P}^{\alpha\beta} P, 
\label{eq:canonical-energy-momentum-fluid}
\\
\Theta^{\alpha\beta}&= \left(\dfrac{e }{c^2}-\frac{\chi}{2}s^2\right)u^\alpha u^\beta+ \mathcal{P}^{\alpha \beta} P -\frac{\chi c^2}{4}s^2g^{\alpha\beta}
\nonumber \\
&+2\left(\dfrac{u_\mu u^\gamma}{c^2}-\delta^\gamma_\mu\right)\hat{\nabla}_\gamma\left[s^{\mu(\alpha}u^{\beta)}\right]. \label{eq:T_tilde_fluid}
\end{align}
\end{subequations}

The general form of the four-momentum density is 
\begin{align}
    p^\alpha = (-p_\mu u^\mu)\dfrac{ u^\alpha}{c^2} + \ell^\alpha , 
\end{align}
where $\ell^\alpha$ is a spacelike vector orthogonal to $u^\alpha$. An  inspection of Eq. (\ref{eq:canonical-energy-momentum-fluid}) suggests $\ell^\alpha =-2a_\sigma s^{\sigma \alpha}/c^2$ and hence
\begin{equation}\label{eq:four-momentum-density}
    p^\alpha= \dfrac{1}{c^2} \left( e u^\alpha - 2 a_\sigma s^{\sigma \alpha}\right),
\end{equation}
upon exploiting  the  relation $ e =  -p_\alpha u^\alpha$ \cite{Obukhov1987}.

By substituting  Eq. \eqref{eq:four-momentum-density} in   Eq. (\ref{eq:canonical-energy-momentum-fluid}),  we  recover the usual form of the  canonical energy-momentum tensor for the Weyssenhoff fluid  \cite{Hehl1976_fundations},  i.e.,   
\begin{equation}
\mathbb{T}^{\alpha \beta} =  p^\alpha u^\beta +\mathcal{P}^{\alpha\beta} P.
\label{eq:canonical-energy-momentum-fluid_2}
\end{equation}

The internal structure of the fluid is described in terms of the rest-mass density $\rho$ and the internal energy density $\varepsilon$.   The total energy density $e$ can be  written as (cf. Eq. \eqref{eq:hydrodynamics-total-en-dens-general-express})
\begin{align}
e &\equiv e(\rho,\mathfrak{s},s_{\mu \nu})= \rho c^2+\varepsilon(\rho, \mathfrak{s},s_{\mu \nu}),
\end{align}
where $\mathfrak{s}$  denotes the specific  entropy (i.e., the entropy per unit mass). The functional dependence of  $\varepsilon$ partially resembles  that of classical physics, apart from the presence of  $s_{\mu\nu}$, which  embodies the new contribution due to quantum mechanical effects and  is specific of the EC theory. 

We suppose that the fluid is \emph{adiabatic} and that the rest mass of the system  is conserved. The first hypothesis can be written as  \cite{Lichnerowicz1967,Rezzolla2013}
\begin{equation}\label{eq:adiabatic-fluid}
\dot{\mathfrak{s}}=0,
\end{equation}
whereas the second one is represented by Eq. (\ref{eq:CONSEN}).\\

The translational equations pertaining to the fluid dynamics can be obtained from Eq. \eqref{eq:EUL}, along with Eqs. \eqref{eq:mu_tensor},  \eqref{eq:spin-tensor-fluid}, \eqref{eq:torsion&contorsion&tau_simplified}, and  \eqref{eq:Phi_fluid}. As pointed out before, the projection along the fluid four-velocity $u^\mu$ gives the energy-balance law (or energy-conservation equation), which upon exploiting Eq. (\ref{eq:CONSEN}) reads as
\begin{equation} \label{eq:energy-conservation-equation-fluid}
   \dot{\varepsilon} + \left(\varepsilon + P \right) \hat{\nabla}_\nu u^\nu=0,
\end{equation}
whereas the projection onto the hypersurface orthogonal to $u^\mu$ gives the  Euler equation, which after some algebra can be written as 
\begin{equation} \label{eq:translational_fluid_equation_2}
\begin{split}
  &  \mathcal{P}^\nu_\mu \partial_\nu P + \dfrac{1}{c^2} \left(P+ e \right) a_\mu - \dfrac{2}{c^2} \hat{\nabla}_\nu \left( u^\nu a^\rho s_{\rho \mu} \right) \\
  &+\chi a^\lambda s_{\lambda \rho} s_{\mu}^{\phantom{\mu}\rho}= - s_{\nu \rho} u^\sigma R_{\mu \sigma}^{\phantom{\mu \sigma}\nu \rho}.
\end{split}  
\end{equation}
We note that, for $s_{\mu \nu}=0$,  Eq. \eqref{eq:translational_fluid_equation_2} reduces to the Euler equation of GR  \cite{Rezzolla2013,Poisson-Will2014}.

Bearing in mind Eqs \eqref{eq:TENS}, \eqref{eq:spin-tensor-fluid},  \eqref{eq:torsion&contorsion&tau_simplified}, and \eqref{eq:canonical-energy-momentum-fluid}, we find that the rotational fluid motion is ruled by
\begin{align} \label{eq:rotational_fluid_equation}
    \mathscr{D}s_{\mu \nu}  & = \dfrac{a^\sigma}{c^2} \left(u_\mu   s_{ \sigma \nu}- u_\nu   s_{ \sigma \mu} \right),
\end{align}
where we have exploited Eq. (\ref{eq:four-momentum-density}). Since the tensor $s_{\mu \nu}$ has in general six independent components,   the Frenkel condition \eqref{eq:Frenkel_condition} is crucial to correctly account for the three true dynamical degrees of freedom associated with the spin of a particle. For this reason, Eq.  \eqref{eq:rotational_fluid_equation} gives rise to three  independent equations.

As discussed in Sec. \ref{sec:rel_hyd}, the system of the cardinal hydrodynamic equations can be closed once the needed constitutive equations are provided. Therefore,  the full set of differential equations governing the Weyssenhoff fluid dynamics is (see Eqs.  \eqref{eq:cons-energy} and  \eqref{eq:adiabatic-fluid}--\eqref{eq:rotational_fluid_equation})
\begin{subnumcases}{\label{system of gravity-matter equations}}
\mathscr{D}\rho=0,
 \label{grav-matter-1}
\\
\mathcal{P}^\nu_\mu \partial_\nu P + \dfrac{1}{c^2} \left(P+ e \right) a_\mu - \dfrac{2}{c^2} \hat{\nabla}_\nu \left( u^\nu a^\rho s_{\rho \mu} \right) 
\nonumber \\
\qquad+\chi a^\lambda s_{\lambda \rho} s_{\mu}^{\phantom{\mu}\rho}+ s_{\nu \rho} u^\sigma R_{\mu \sigma}^{\phantom{\mu \sigma}\nu \rho}=0,
\label{grav-matter-2}
\\
\mathscr{D}s_{\mu \nu} - \dfrac{a^\sigma}{c^2} \left(u_\mu   s_{ \sigma \nu}- u_\nu   s_{ \sigma \mu} \right)=0,
\label{grav-matter-3}\\
\dot{\varepsilon} + \left(\varepsilon + P \right) \hat{\nabla}_\nu u^\nu=0,
\label{grav-matter-4}
\\
\dot{\mathfrak{s}}=0,
\label{grav-matter-5}
\\
P=P(\rho,\varepsilon).
\label{grav-matter-6}
\end{subnumcases}

The above system  contains ten independent equations and exactly ten unknowns, i.e., the three components of the fluid velocity $u^\mu$, the rest-mass density  $\rho$, the internal energy density $\varepsilon$, the specific entropy $\mathfrak{s}$, the isotropic pressure $P$, and the three independent components of $s_{\mu \nu}$. It is also worth noting that  it explicitly includes an equation describing the behaviour of  $\mathfrak{s}$ (cf. Eq. \eqref{grav-matter-5}). This is due to the fact that, unlike GR \cite{Rezzolla2013,Poisson-Will2014}, in EC theory Eqs. \eqref{grav-matter-1} and  \eqref{grav-matter-4} do not imply that the fluid evolves adiabatically (see Sec. \ref{Sec:first-thermodyanamic-law}).

In conclusion, the closed self-consistent system of gravity-matter equations is represented by EC field equations   (\ref{eq:Einstein-Cartan_equations})  supplemented by  the set \eqref{system of gravity-matter equations}.

\subsubsection{The first thermodynamic law}
\label{Sec:first-thermodyanamic-law}

In the study of the  hydrodynamics, it is important to investigate the implications of the energy-conservation equation \eqref{grav-matter-4} at  thermodynamic level \cite{Obukhov1987,Rezzolla2013}. However, we note that this analysis does not provide additional constraining equations to the system \eqref{system of gravity-matter equations}, but it allows to better understand how the fluid behaves in EC theory.

The \emph{first thermodynamic law} for a spinning fluid can be written as \cite{Ray1982a,deRitis1983,Amorim(1985),Obukhov1987} 
\begin{equation}\label{eq:first-thermodynamic-law}
    {\rm d} \Pi  =\dfrac{P}{\rho^2} {\rm d}\rho + \theta {\rm d}\mathfrak{s} + \dfrac{1}{2} \dfrac{\omega_{\mu \nu}}{\rho} {\rm d}s^{\mu \nu}, 
\end{equation}
where 
\begin{equation} \label{eq:specific-internal-energy}
    \Pi  \equiv \Pi(\rho,\mathfrak{s},s_{\mu \nu}) \equiv \dfrac{\varepsilon}{\rho},
\end{equation}
is the \emph{specific internal energy}, and
\begin{subequations}
\begin{align}
\left(\dfrac{\partial \Pi}{\partial \rho}\right)_{\mathfrak{s},s^{\mu \nu}} &\equiv \dfrac{P}{\rho^2}, 
\\
\left(\dfrac{\partial \Pi}{\partial \mathfrak{s}}\right)_{\rho,s^{\mu \nu}} &\equiv \theta, 
\\
\left(\dfrac{\partial \Pi}{\partial s^{\mu \nu}}\right)_{\rho,\mathfrak{s}} &\equiv \dfrac{1}{2} \dfrac{\omega_{\mu \nu}}{\rho}, 
\end{align}
\end{subequations}
$\theta$  being the temperature and $\omega_{\mu \nu}$  the thermodynamic variable conjugated to $s^{\mu \nu}$ coinciding with the microscopic angular velocity of the fluid. 

Starting from the continuity equation \eqref{grav-matter-1} and the energy-balance law \eqref{grav-matter-4}, we obtain 
\begin{equation}\label{eq:thermodynamic-relation-3}
 \dot{\Pi} - \dfrac{P}{\rho^2}  \dot{\rho}=0,
\end{equation}
which, once compared with Eq. (\ref{eq:first-thermodynamic-law}), gives
\begin{equation} \label{eq:thermodynamic-relation-1}
    \theta \dot{\mathfrak{s}} + \dfrac{1}{2} \dfrac{\omega^{\mu \nu}}{\rho} \dot{s}_{\mu \nu}=0.
\end{equation}
Bearing in mind Eq. (\ref{grav-matter-5}), it follows that 
\begin{equation} \label{eq:thermodynamic-relation-2}
    \omega^{\mu \nu} \dot{s}_{\mu \nu}=0.
\end{equation}
At this point, some remarks are in order. First of all, as pointed out before, it is known that in  GR a  perfect fluid evolves adiabatically \cite{Rezzolla2013,Poisson-Will2014}. This result can be proved by following the same procedure as the one we have adopted in this section, i.e., by exploiting the fluid continuity equation jointly with the energy-balance law. On the contrary, in EC theory the same calculations do not imply the adiabatic-flow hypothesis as long as the spin contribution in Eq. \eqref{eq:thermodynamic-relation-1} does not vanish. This shows that in EC hydrodynamics the adiabatic condition must be explicitly imposed   via  Eq. \eqref{grav-matter-5}, which in fact has been employed to obtain the final relation  (\ref{eq:thermodynamic-relation-2}). 

\subsection{Post-Newtonian analysis of the Weyssenhoff fluid dynamics} 
\label{Sec:PN-Hydrodynamics}

In this section, we perform the PN investigation of the Weyssenhoff fluid dynamics, which is useful both for the calculations carried out in Sec. \ref{sec:Point-particle-limit} and, in general, for several other applications. In Sec. \ref{sec:PN-Comp-program}, we outline the  PN computational scheme. In Sec. \ref{Sec:Preliminary-PN-expansions}, we present the PN expansion of some basic quantities which are then exploited to compute the 0PN and 1PN orders of Eq. \eqref{system of gravity-matter equations} in Secs. \ref{sec:0PN-eq} and \ref{sec:1PN-eq}, respectively.

\subsubsection{The computational program}
\label{sec:PN-Comp-program}

The EC hydrodynamics, although equipped with the simplifying constraint $S_{\alpha\beta}{}^\beta=0$, gives rise to a set of coupled nonlinear differential equations (cf. Eq. \eqref{system of general matter equations}). Also in the more specific case, where we consider the Weyssenhoff fluid endowed with the Frenkel condition \eqref{eq:Frenkel_condition}, the nonlinear nature and the differential structure still endure (cf. Eq. \eqref{system of gravity-matter equations}), and we are, in general, not able to solve the mathematical problem exactly. An analytical solution may be determined by imposing symmetry requirements, like time independence and spatial isotropy (as it occurs, for example, in GR for the Schwarzschild metric). However, this approach reveals to be very limited in its range of applications, where time variability and spatial anisotropies occur frequently. 

Since the EC gravity framework is geometrically more tangled than  GR theory, our goal is therefore not to search for further exact solutions, but rather to take advantage of a \emph{comprehensive and solid approximation method}, which goes beyond the symmetries and the particular functional form of the starting problem. The mathematical formulation of the GW generation problem naturally suggests us to resort to the \emph{perturbation theory} (see Refs. \cite{Weinberg1972,Bender1999,Holmes2012}, for further details and examples). This approach permits to determine a solution under the form of a power series of a small parameter $\epsilon$ (represented, in our case, by $\epsilon=1/c$), where expressions of higher powers of $\epsilon$ become smaller as the  order increases. The \emph{approximate perturbed solution} is obtained by truncating such a series at a certain established order, suggested either by the observational sensitivity of the phenomenon under investigation or, from a  theoretical point of view, by the computational complexity (i.e., gradually increasing the PN order calculations).

The strategy to build up the solution via the PN formalism  relies  on  expanding  all  the  involved  quantities, and then extracting the coefficients of the expansions which are useful to construct the differential equations at each PN level. In this way, we obtain a series of \emph{PN problems}, where the lowest order permits  obtaining the classical equations with eventual spin corrections, and the  higher terms provide the successive perturbative corrections. The method is therefore iteratively based on first determining (either analytically or numerically) the solution of the 0PN problem, and then proceeding \emph{order by order} to solve (either analytically or numerically) the successive PN problems. Indeed, to compute the $N$th PN order, we need to use the  parameters determined in the previous iterations. Finally, gathering all these quantities together we can provide an approximate solution to the original problem up to the desired PN level.

It is important to remark that we  apply this computational program in the derivation of the 0PN and 1PN orders of Eq. \eqref{system of gravity-matter equations}  without solving them. As we will see,  the obtained results  will be useful in Sec.  \ref{sec:Point-particle-limit}.

\subsubsection{Preliminary:  expansions of basic quantities}
\label{Sec:Preliminary-PN-expansions}

Given the harmonic coordinates $x^\mu = (ct,\boldsymbol{x})$ (cf. Eq. \eqref{eq:prob-EC2}), the 1PN series of the metric tensor, expressed in terms of the retarded potentials $V$ and $V_i$,   reads as \cite{Paper1}
\begin{subequations}
\label{eq:1PN_metric_torsion}
\begin{align} 
g_{00} &= - {\rm e}^{-2V/c^2} + {\rm O}(c^{-6}),  \\
g_{0i} &=  -\dfrac{4}{c^3} V_i + {\rm O}(c^{-5}), \\
g_{ij} &= \delta_{ij} \left(1 + \dfrac{2}{c^2}V\right)+ {\rm O}(c^{-4}), 
\end{align}
\end{subequations}
where  (cf. Eq. \eqref{eq:def_sigma-sigma_i})
\begin{subequations}
\label{eq:potential V and V-i}
\begin{align}
V\left(t,\boldsymbol{x} \right) &=  G \int \dfrac{{\rm d}^3 \boldsymbol{y}}{\vert \boldsymbol{x}- \boldsymbol{y}\vert} \sigma \left(t- \vert \boldsymbol{x}- \boldsymbol{y}\vert/c, \boldsymbol{y}\right),
\label{eq:definition-of-retarded-potential-V}
\\
V_i \left(t,\boldsymbol{x} \right) &= G \int \dfrac{{\rm d}^3 \boldsymbol{y} }{\vert \boldsymbol{x}- \boldsymbol{y}\vert} \sigma_i \left(t- \vert \boldsymbol{x}- \boldsymbol{y}\vert/c. \boldsymbol{y}\right).
\label{eq:potential-V_i-def}
\end{align}
\end{subequations}
The expansion of Eq. (\ref{eq:potential V and V-i}) for small retardation effects gives 
\begin{subequations}
\label{eq:potentials_V,U,X,Vi,Ui}
    \begin{align} 
    V &= U + \dfrac{1}{2c^2} \partial^2_t X + {\rm O}\left(c^{-3}\right), \label{eq:V-in-terms-U-and-X}
    \\
  V_i &= U_i + {\rm O}\left(c^{-2}\right),  
 \end{align}
\end{subequations}
the potentials $U$, $U_i$,  and the superpotential $X$ being given by 
\begin{subequations}
    \begin{align}
     U\left(t,\boldsymbol{x} \right) &= G \int \dfrac{{\rm d}^3 \boldsymbol{y}}{\vert \boldsymbol{x}- \boldsymbol{y}\vert} \sigma \left(t, \boldsymbol{y}\right), 
     \label{eq:U-potential-def}
     \\
     U_i\left(t,\boldsymbol{x} \right) &= G \int \dfrac{{\rm d}^3 \boldsymbol{y}}{\vert \boldsymbol{x}- \boldsymbol{y}\vert} \sigma_i \left(t, \boldsymbol{y}\right),
     \\
      X(t,\boldsymbol{x})&= G \int {\rm d}^3 \boldsymbol{y}\, \vert \boldsymbol{x}- \boldsymbol{y}\vert \, \sigma \left(t, \boldsymbol{y}\right),
\end{align}
\end{subequations}
respectively. It should be  noted that the  term of order $c^{-3}$ in Eq. \eqref{eq:V-in-terms-U-and-X} is a function of time only. 

Stating from Eq. \eqref{eq:1PN_metric_torsion},  one easily obtains the PN structure of the Christoffel symbols (see Ref. \cite{Weinberg1972}, for details.) 

The spin angular momentum tensor  $\tau_{\lambda}^{\phantom{\lambda}\mu \nu}$  admits, in general, the following PN series  \cite{Paper1}:
\begin{align}
 \tau_{0}^{\phantom{\lambda}i 0} &={}^{(0)}\tau_{0}^{\phantom{\lambda}i 0}+ {}^{(2)}\tau_{0}^{\phantom{\lambda}i 0} + {\rm O}\left(c^{-2}\right), \nonumber\\
 \tau_{0}^{\phantom{\lambda}i j} &={}^{(1)}\tau_{0}^{\phantom{\lambda}i j}+ {}^{(3)}\tau_{0}^{\phantom{\lambda}i j} + {\rm O}\left(c^{-3}\right), \nonumber\\
 \tau_{i}^{\phantom{\lambda} j 0} &={}^{(1)}\tau_{i}^{\phantom{\lambda}j 0}+ {}^{(3)}\tau_{i}^{\phantom{\lambda}j 0} + {\rm O}\left(c^{-3}\right), \nonumber\\
 \tau_{i}^{\phantom{\lambda} j k} &={}^{(2)}\tau_{i}^{\phantom{\lambda}j k}+ {}^{(4)}\tau_{i}^{\phantom{\lambda}j k} + {\rm O}\left(c^{-4}\right),      
 \label{eq:tau_PN_expansion}
\end{align}
where $ {}^{(n)}\tau_{\lambda}^{\phantom{\lambda}\mu \nu}  \sim \dfrac{\bar{M}}{\bar{d}^2} \dfrac{\bar{v}^n}{c^{n-2}}$ ($\bar{v}$ and $\bar{M}$  being some typical  internal velocity and  mass of the source, respectively). Therefore, the tensor 
(\ref{Eq:S_alphabeta}) has the PN structure
\begin{align}
 \label{eq:S_expansion}
       & \mathcal{S}^{00}={}^{(0)}\mathcal{S}^{00} + {}^{(1)}\mathcal{S}^{00} +{}^{(2)}\mathcal{S}^{00} + {\rm O}\left(c^{-2}\right), \nonumber\\
       & \mathcal{S}^{0i}={}^{(0.5)}\mathcal{S}^{0i} + {}^{(1.5)}\mathcal{S}^{0i} +{}^{(2.5)}\mathcal{S}^{0i} + {\rm O}\left(c^{-3}\right), \nonumber\\  
            & \mathcal{S}^{ij}={}^{(0)}\mathcal{S}^{ij} + {}^{(1)}\mathcal{S}^{ij} +{}^{(2)}\mathcal{S}^{ij}+ {\rm O}\left(c^{-2}\right),
\end{align}
with ${}^{(n)}\mathcal{S}^{\mu \nu}  \sim  \dfrac{\bar{M}^2}{\bar{d}^4} \dfrac{\bar{v}^{2n}}{c^{2n-4}}$.

Starting from  Eq. \eqref{eq:tau_PN_expansion} along with the PN expression of the Christoffel symbols, it is possible to derive the PN form of the   torsion and contortion tensors  \eqref{eq:torsion&contorsion&tau_simplified},   the connection coefficients  \eqref{eq:Affine_Connection_1}, and the  Riemann tensor \eqref{eq:EC-Riemann-tensor}.  The  components of the the  Riemann tensor needed at 1PN level are reported in \ref{sec:Riemann}.

Bearing in mind the definition of the spin angular momentum of a
test particle in EC theory (see Eq. (40) of Ref. \cite{Paper1}),  for the spin density tensor we find
\begin{equation}
\label{eq:spin_density_PN_expansion}
\begin{split}
        s_{0i} &= {}^{(0)}s_{0i} +{}^{(2)}s_{0i} +{\rm O}(c^{-3}), \\
    s_{ij} &= {}^{(1)}s_{ij} + {}^{(3)}s_{ij} +{\rm O}(c^{-4}),
\end{split}
\end{equation}
where  ${}^{(n)}s_{\mu \nu}$ indicates a factor going like $\dfrac{\bar{M} \bar{v}^n}{\bar{d}^2 c^{n-1}}$. If we write the  fluid four-velocity  equivalently  as 
\begin{equation}\label{4-velocity_2}
    u^\mu = \gamma \left(c,\boldsymbol{v}\right),
\end{equation}
(with $\gamma \equiv \tfrac{u^0}{c}$ and 
$\boldsymbol{v} \equiv \tfrac{{\rm d}\boldsymbol{x}}{{\rm d}t}$) its PN expansion can be easily constructed by means of Eq.  \eqref{eq:1PN_metric_torsion} and the normalization condition. Therefore,    the  Frenkel condition (\ref{eq:Frenkel_condition}) leads to 
\begin{subequations}
\label{eq:PN-exp-Frenkel-cond}
\begin{align}
&    {}^{(0)}s_{i0} \left({}^{(1)}u^i + {}^{(3)}u^i\right)\notag\\
&+   {}^{(2)}s_{i0} \, {}^{(1)}u^i + {\rm O}(c^{-3})=0, \label{eq:temp_FC}
\\
&    {}^{(0)}s_{j0} \left({}^{(0)}u^0 + {}^{(2)}u^0\right) +   {}^{(2)}s_{j0}\,{}^{(0)}u^0  
\nonumber \\
&+ {}^{(1)}s_{ji} \,{}^{(1)}u^i+ {\rm O}(c^{-2})=0,
\label{eq:spaz_FC}
\end{align}
\end{subequations}
where ${}^{(n)}u^\mu$ denotes a term of order $\dfrac{\bar{v}^n}{c^{n-1}}$.
We note that Eq. \eqref{eq:temp_FC} results from   a linear combination of Eq. \eqref{eq:spaz_FC} with  coefficients  $v^j/c$ (recalling that ${}^{(n)}u^j=\tfrac{{}^{(n-1)}u^0}{c} v^j$ with $n=1,3$). This implies that the latter gives rise to  three  independent relations which can be used to gauge away the  components $s_{0i}$ of the spin density tensor at  different PN orders in the equations of motion (cf. Sec. \ref{sec:hystory} and the discussion below Eq. (\ref{eq:rotational_fluid_equation})). Indeed, from Eq. \eqref{eq:spaz_FC} we obtain to leading order and to next-to-leading order, respectively, 
\begin{subequations}
\label{eq:FC_leading_next_to_leading_1}
\begin{align}
& {}^{(0)}s_{0i} =\rm{O}(c^{-1}),
\\
&{}^{(0)}s_{0i} \left[1+ \frac{1}{c^2}\left(V+\frac{v^2}{2}\right) \right]+{}^{(2)}s_{0i}={}^{(1)}s_{ik} \dfrac{v^k}{c}+\rm{O}(c^{-3}),
\end{align}
\end{subequations}
or, equivalently, 
\begin{subequations}
\label{eq:FC_leading_next_to_leading_2}
\begin{align}
{}^{(0)}s_{0i}&=0,
\label{eq:Frenkel_condition_leading_2a}
\\
{}^{(2)}s_{0i}&=\frac{{}^{(1)}s_{ik}v^k}{c}.
\label{eq:Frenkel_condition_leading_2b}
\end{align}
\end{subequations}
From the above equations it is clear that no components $s_{0i}$ are present at leading order, whereas they can be written in terms of ${}^{(1)}s_{ik}$ at next-to-leading order. Furthermore, we see that Eq. \eqref{eq:temp_FC} simply states that the three-vectors $s_{0i}$ and $u^i$ are perpendicular, but gives no information about the form of  $s_{0i}$. On the contrary, Eq. \eqref{eq:spaz_FC} allows to write explicitly, at each PN level, an expression for $s_{0i}$ which turns out to be both orthogonal to $u^i$ and  vanishing in the fluid rest frame.

For our forthcoming calculations, it is useful to express Eq. (\ref{grav-matter-1}) in an equivalent form. Indeed,  bearing in mind Eq. \eqref{4-velocity_2}, and upon introducing the rescaled mass density $\rho^{\star} \equiv \gamma \sqrt{-g} \rho$ \cite{Poisson-Will2014}, Eq.  (\ref{grav-matter-1}) yields the (exact) equation 
\begin{equation} \label{eq:cons-mass-with-rho-star}
    \partial_t \rho^{\star} + \partial_j \left(\rho^{\star} v^j\right)=0,
\end{equation}
the relation between $\rho^{\star}$ and the proper rest-mass density $\rho$ being
\begin{equation}\label{eq:rho-star-in-terms-of-rho}
    \rho^\star = \rho \left[ 1+\dfrac{1}{c^2}\left(\dfrac{1}{2} v^2 +3V\right) + {\rm O}(c^{-4}) \right].
\end{equation}

The PN expansion of the first thermodynamic law \eqref{eq:first-thermodynamic-law} reads as (cf. Eq. \eqref{eq:specific-internal-energy})
\begin{align}
    \rho^\star \dfrac{{\rm d}\varepsilon}{{\rm d}t} &=\left(\varepsilon+P\right)\dfrac{{\rm d}\rho^\star}{{\rm d}t} + \left(\rho^\star \right)^2 \theta \dfrac{{\rm d}\mathfrak{s}}{{\rm d}t} + \dfrac{1}{2} \rho^\star \omega_{\alpha \beta}\dfrac{{\rm d}s^{\alpha \beta}}{{\rm d}t} 
    \nonumber \\
   &+ {\rm O}\left(c^{-2}\right),
\end{align}
where
\begin{equation}
\dfrac{{\rm d}}{{\rm d}t} f(t,\boldsymbol{x})= \partial_t f + v^k \partial_k f,
 \end{equation}
and we have exploited Eq. \eqref{eq:rho-star-in-terms-of-rho}, the relations $\omega_{0i} = {\rm O}\left(c^{-1}\right)$ and  $\omega_{ij} = {\rm O}\left(c^0\right)$ (see Refs. \cite{Obukhov1987,Li2014} for details), and we have supposed that both $\theta$ and $\mathfrak{s}$ are $\rm{O}(c^0)$ quantities.

\subsubsection{Equations of motion: 0PN expansion}
\label{sec:0PN-eq}

The results contained in  Sec. \ref{Sec:Preliminary-PN-expansions} allow us to  study  the   
 hydrodynamic equations \eqref{system of gravity-matter equations} within the PN formalism.  In our investigation, the rest-mass conservation equation \eqref{grav-matter-1} will be replaced by its equivalent expression (\ref{eq:cons-mass-with-rho-star}) and will not  be expanded. Furthermore, by means  of Eq. \eqref{4-velocity_2},  Eq. \eqref{grav-matter-5} assumes the exact form 
 \begin{equation}
     \dfrac{{\rm d}\mathfrak{s}}{{\rm d}t}=0,
 \end{equation}
and the constitutive equation \eqref{grav-matter-6} will be simply written in terms of $\rho^\star$ (cf. Eq. \eqref{eq:rho-star-in-terms-of-rho}). Therefore,  by computing the 0PN expansion of the remaining equations of the system \eqref{system of gravity-matter equations} and by exploiting the Frenkel condition \eqref{eq:Frenkel_condition_leading_2a},  we end up, after a lengthy calculation, with the following system pertaining to the dynamics of a Weyssenhoff fluid with  0PN accuracy:
\begin{subnumcases}{\label{system of gravity-matter equations-0PN}}
\partial_t \rho^\star +\partial_j(v^j \rho^\star)=0,
\label{eq:0PN-eq1}
\\
\rho^\star \left(\dfrac{{\rm d}v^i}{{\rm d}t} - \partial_i U \right) + \partial_i P={\rm O}\left(c^{-2}\right),
\label{eq:0PN-eq2}
\\
\dfrac{{\rm d}}{{\rm d}t} \left({}^{(1)}s_{ij}\right)+{}^{(1)}s_{ij}\partial_k v^k={\rm O}\left(c^{-2}\right),
\label{eq:0PN-eq3}
\\
\dfrac{{\rm d}\varepsilon}{{\rm d}t}+\left(\varepsilon + P\right)\partial_k v^k={\rm O}\left(c^{-2}\right),
\label{eq:0PN-eq4}
\\
\dfrac{{\rm d}\mathfrak{s}}{{\rm d}t}=0,
\label{eq:0PN-eq5}
\\
P=P(\rho^\star,\varepsilon).
\label{eq:0PN-eq6}
\end{subnumcases}

Equation \eqref{eq:0PN-eq2} originates from the 0PN expansion of Eq.  (\ref{grav-matter-2}) and  represents the Euler equation of the Newtonian theory. Equation \eqref{eq:0PN-eq3} is the leading-order piece of Eq. (\ref{grav-matter-3}). \emph{At this level, only the derivative $\mathscr{D}s_{i j}$ gives a contribution and we end up with  a homogeneous  continuity equation for} ${}^{(1)}s_{ij}$. Finally, Eq.  \eqref{eq:0PN-eq4} is the 0PN-accurate energy-conservation equation  (\ref{grav-matter-4}).

\subsubsection{Equations of motion: 1PN expansion}
\label{sec:1PN-eq}

In this section, we present the 1PN expressions of Eqs. (\ref{grav-matter-2})--(\ref{grav-matter-4}). 

The 1PN Euler equation (\ref{grav-matter-2}) reads as   
\begin{align}
& \rho^\star \left(\dfrac{{\rm d}v^i}{{\rm d}t} - \partial_i U \right) + \partial_i P
+\dfrac{v^i}{c^2}\dfrac{{\rm d}P}{{\rm d}t} -\dfrac{1}{2c^2} \rho^\star \partial_i \partial^2_t X
\notag \\
&+\dfrac{1}{c^2}\left(\dfrac{{\rm d}v^i}{{\rm dt}} -\partial_i U\right)\left[P+\varepsilon-\rho^\star\left(\frac{v^2}{2}+3U\right)\right]
\notag\\
&+\dfrac{\rho^\star}{c^2}\left\{\left(\frac{v^2}{2}+3U\right)\dfrac{{\rm d}v^i}{{\rm d}t}+ 2 v^i \dfrac{{\rm d}U}{{\rm d}t} - 4 \dfrac{{\rm d}U_i}{{\rm d}t} \right.
\notag\\
&+ \left. \dfrac{{\rm d}}{{\rm d}t}\left[v^i\left(U+\dfrac{v^2}{2} \right)\right]-2 v^2 \partial_i U  + 4 v^l \partial_i U_l \right\}
\notag\\
&-\dfrac{2}{c^2}\left\{\dfrac{{\rm d}}{{\rm d}t}\left[{}^{(1)}s_{ki}\left(\dfrac{{\rm d}v^k}{{\rm dt}} -\partial_k U\right)\right]\right.
\notag\\
&\left.+\left(\dfrac{{\rm d}v^k}{{\rm dt}} -\partial_k U\right){}^{(1)}s_{ki}\partial_lv^l\right\}+\dfrac{2}{c^2}\,{}^{(1)}s_{jk}\Biggr{[}-v^k\partial_i\partial_j U
\notag\\
&+v^l\left(\delta_{i[k}\partial_{j]}\partial_l+\delta_{l[j}\partial_{k]}\partial_i\right)U+\frac{\chi\,c^4}{2}\partial_{[k} {}^{(1)}s_{i|j]}\notag\\
&+2\partial_i \partial_{[j} U_{k]}+\delta_{i[k}\partial_{j]}\partial_t U\Biggr{]}={\rm O}\left(c^{-4}\right),
\label{eq:1PN-Euler-equation-explicit}
\end{align}
whereas at 1PN order the spin equation (\ref{grav-matter-3}) becomes
\begin{align}
&\dfrac{{\rm d}}{{\rm d}t} \left({}^{(1)}s_{ij}\right)+{}^{(1)}s_{ij}\partial_k v^k 
\nonumber \\
&+\dfrac{{\rm d}}{{\rm d}t} \left[\dfrac{{}^{(1)}s_{ij}}{c^2} \left(\dfrac{v^2}{2}+U\right)
  + {}^{(3)}s_{ij}\right]   
  \nonumber \\
  &+ \left[\dfrac{{}^{(1)}s_{ij}}{c^2} \left(\dfrac{v^2}{2}+U\right) +{}^{(3)}s_{ij}\right]  \partial_k v^k 
  \nonumber \\
 &+ \dfrac{1}{c^2}   \dfrac{{\rm d}v^k}{{\rm d}t} \left[ v^j \, {}^{(1)}s_{ki}-v^i \, {}^{(1)}s_{kj} \right] 
 \nonumber \\
 &+\dfrac{2}{c^2} \Biggl[ {}^{(1)}s_{ki} \left(\partial_k U_j -\partial_j U_k + v^k \partial_j U -\dfrac{v^j}{2} \partial_k U\right)
 \nonumber \\
 &- {}^{(1)}s_{kj} \left(\partial_k U_i -\partial_i U_k + v^k \partial_i U -\dfrac{v^i}{2} \partial_k U\right)\Biggr]
 \nonumber \\
 &+ \dfrac{1}{c^2} \left(v^i \,{}^{(1)}s_{lj} \partial_l U - v^j \,{}^{(1)}s_{li} \partial_l U  \right)={\rm O}\left(c^{-4}\right).
 \label{eq:1PN-spin-equation-explicit}
\end{align}
In deriving Eqs. \eqref{eq:1PN-Euler-equation-explicit} and \eqref{eq:1PN-spin-equation-explicit}, we have exploited Eqs. (\ref{eq:potentials_V,U,X,Vi,Ui}) and \eqref{eq:FC_leading_next_to_leading_2}. 

The energy-balance law (\ref{grav-matter-4}) yields, at 1PN order,
\begin{align}
&\dfrac{{\rm d}\varepsilon}{{\rm d}t} +\left(\varepsilon+P\right)\partial_j v^j +\dfrac{1}{c^2} \left(\dfrac{v^2}{2}+U\right) \dfrac{{\rm d}\varepsilon}{{\rm d}t} 
\nonumber \\
&+\dfrac{\left(\varepsilon + P\right)}{c^2}\Biggl[   \dfrac{{\rm d}}{{\rm d}t} \left(\dfrac{v^2}{2}+3U\right)+ \left(\dfrac{v^2}{2}+U\right) \partial_k v^k \Biggr]
\nonumber \\
&={\rm O}\left(c^{-4}\right),
\label{eq:1PN-energy-equation-explicit}
\end{align}
where we have exploited Eq. (\ref{eq:V-in-terms-U-and-X}). 

Therefore, the Weyssenhoff fluid is described, at 1PN order, by the following system:
\begin{subnumcases}{\label{system of gravity-matter equations-1PN}}
\partial_t \rho^\star +\partial_j(v^j \rho^\star)=0,
\label{eq:1PN-eq1}
\\
\mbox{Equation \eqref{eq:1PN-Euler-equation-explicit}},
\label{eq:1PN-eq2}
\\
\mbox{Equation \eqref{eq:1PN-spin-equation-explicit}},
\label{eq:1PN-eq3}
\\
\mbox{Equation \eqref{eq:1PN-energy-equation-explicit}},
\label{eq:1PN-eq4}
\\
\dfrac{{\rm d}\mathfrak{s}}{{\rm d}t}=0,
\label{eq:1PN-eq5}
\\
P=P\left(\rho^\star, \varepsilon\right).
\label{eq:1PN-eq6}
\end{subnumcases}

\section{The point-particle limit of the Weyssenhoff fluid}
\label{sec:Point-particle-limit}

The final outcome of the PN treatment contained  in Sec. \ref{Sec:Application-to-Weyssenhoff-fluid} is represented by  Eqs. \eqref{system of gravity-matter equations-0PN} and \eqref{system of gravity-matter equations-1PN}, which completely determine the behaviour of the  Weyssenhoff fluid at 0PN and 1PN level, respectively. In general, these two sets comprise both partial and integro-differential equations, whose resolution, either analytically or via numerical means, turns out to be highly challenging. This makes, as a consequence, the evaluation of the radiative moments parametrizing the asymptotic waveform and the  radiated power (cf. Sec. \ref{sec:GW_PR}) rather demanding.   

A widely used approach in the literature which offers a way around this issue consists in employing the so-called \emph{point-particle procedure}, where the fluid configuration is supposed to be described by a collection of separated point-like components, usually refereed to as \qm{bodies}. Within this pattern, the equations underlying the fluid dynamics become less complex as they are turned into \emph{ordinary differential equations}. For these reasons, in Sec. \ref{Sec:PP-limit-subsec}, we introduce the point-particle procedure and work out the 1PN-accurate expressions of the radiative moments. As we will see, the results of Sec. \ref{Sec:Application-to-Weyssenhoff-fluid} will be crucial, since they allow us to go from the fine-grained description to the coarse-grained picture of the Weyssenhoff fluid. Finally, in Sec. \ref{Sec:Binary-System}, we deal with the special case of binary systems, which are known to be the main candidates of GW events in high-energy astrophysics.

\subsection{The point-particle procedure}
\label{Sec:PP-limit-subsec}

The evaluation of the 1PN-accurate total power of emission \eqref{eq:power-radiated-1PN} requires
the knowledge of the 1PN radiative  mass quadrupole moment $U_{ij}$ as well as  the 0PN mass octupole $U_{ijk}$ and  current quadrupole $V_{ij}$; on the other hand,  the 1PN asymptotic waveform \eqref{eq:gravitational_wave_amplitude} can be worked out once  the 0PN mass $2^4$-pole $U_{ijkl}$ and current octupole $V_{ijk}$ are also known. Having solved the 1PN GW generation problem \cite{Paper1}, we know that these  are related to the source multipole moments via Eqs. \eqref{eq:radiative-source_1PN} and \eqref{eq:I_L and J_L-torsion}. Therefore, in order to compute the point-particle limit of the required radiative moments, we first need to determine their fine-grained  expression, i.e., the form assumed  in the case of a continuous distribution of  matter  represented by the Weyssenhoff fluid, see Sec. \ref{sec:preamble}. Then, after some premises outlined in Secs. \ref{sec:PP-limit-preliminary} and \ref{sec:coarse-grained-descr}, the point-particle limit is performed in Sec. \ref{sec:PP-limit-results}.

\subsubsection{The fine-grained form of the radiative moments}
\label{sec:preamble}  

In this section, we calculate the fine-grained  expression of the radiative moments  occurring  in
the   total power of emission   and the  wave amplitude for the specific case of the   Weyssenhoff fluid. This entails, first of all,  the computation of   the PN expansion of metric stress-energy tensor (\ref{eq:T_alpha_beta_fluid}) and the tensor (\ref{eq:S-tensor-fluid}). With the help of the results of Sec. \ref{Sec:Preliminary-PN-expansions},  the former yields
\begin{subequations}
\label{eq:PN-expansion-T-alpha-beta}
\begin{align}
 {}^{(0)} T^{00} &= \rho c^2,   
 \\
 {}^{(2)} T^{00} &= \varepsilon +\rho(2U+v^2)-2\partial_i\left({}^{(1)}s_{ik} v^k\right), 
 \\
{}^{(1)} T^{0i} &= c\left[\rho v^i-\partial_k\left({}^{(1)}s_{ki}\right)\right], 
\\
{}^{(2)} T^{ij} &= \rho v^iv^j+\delta^{ij}P-2\partial_k\left({}^{(1)}s_{k(i}v^{j)}\right),
\end{align}
\end{subequations}
whereas the latter
\begin{equation}
\label{eq:PN-expansion-S-alpha-beta}
{}^{(0)}\mathcal{S}^{00}={}^{(0.5)}\mathcal{S}^{0i}={}^{(0)}\mathcal{S}^{ij}=0,    
\end{equation}
where we have exploited Eqs. (\ref{eq:V-in-terms-U-and-X}) and \eqref{eq:FC_leading_next_to_leading_2}, and ${}^{(n)}{T}{_{\mu \nu}}$   stands for the contributions in ${T}{_{\mu \nu}}$ of order $\dfrac{\bar{M} \bar{v}^n}{\bar{d}^3 c^{n-2}}$.   

By means of Eqs. (\ref{eq:rho-star-in-terms-of-rho}), \eqref{eq:PN-expansion-T-alpha-beta}, and \eqref{eq:PN-expansion-S-alpha-beta}, we have
\begin{subequations}
\label{eq:sigma-sigmai-sigmaii-Weyseenhoff}
\begin{align}
\sigma&=\rho^{\star \star} + \rho_{{\rm v}}  -\dfrac{4}{c^2}\partial_k\left(s_{kl}v^l\right) + {\rm O}\left( c^{-4}\right),
\label{Eq:sigma-EC-theory}
\\
\sigma_i&=\rho^\star v^i-\partial_k s_{ki} +{\rm O}\left( c^{-2}\right),
\\
\sigma_{ii}&=\rho^\star\left( v^2+ \dfrac{3P}{\rho^\star}\right)-2\partial_k\left(s_{kl}v^l\right) + {\rm O}\left( c^{-2}\right),\label{Eq:sigma-EC-theory-ii}
\end{align}
\end{subequations}
where $\sigma$ and $\sigma_i$ have been defined in Eq. (\ref{eq:def_sigma-sigma_i}), while  $ \sigma_{ij} \equiv \Theta^{ij}$ \cite{Paper1}, and, inspired by Ref. \cite{Blanchet-Schafer1989},
\begin{align}
\rho^{\star \star} & \equiv \rho^\star \left[1+\dfrac{1}{c^2} \left(\dfrac{v^2}{2}+ \Pi -\dfrac{U}{2}\right)\right],   
\\
\rho_{\rm v}& \equiv \dfrac{1}{c^2} \rho^\star \left(v^2 -\dfrac{U}{2}+ \dfrac{3P}{\rho^\star}\right),
\end{align}
with (see Eq. \eqref{eq:U-potential-def})
\begin{align} \label{eq:potential-U-Weyssenhoff-fluid}
U\left(t,\boldsymbol{x} \right) &= G \int \dfrac{{\rm d}^3 \boldsymbol{x}^\prime}{\vert \boldsymbol{x}- \boldsymbol{x}^\prime\vert} \rho^\star \left(t,\boldsymbol{x}^\prime\right) +{\rm O}\left(c^{-2}\right).
\end{align}
It should be noticed that the spin contributions in Eq. \eqref{eq:sigma-sigmai-sigmaii-Weyseenhoff} appear both explicitly, via the terms involving $s_{ij}$, and implicitly, through, e.g.,  the specific internal energy  $\Pi$ (see Eq. \eqref{eq:specific-internal-energy}).

For our forthcoming analysis, it is useful to  define a new set of STF mass-type radiative moments $I_L^{{\rm rad}}$ and current-type radiative moments  $J_L^{{\rm rad}}$ according to
\begin{align}
U_L &\equiv \overset{(l)}{I}{}_{L}^{{\rm rad}},\qquad 
V_L \equiv\overset{(l)}{J}{}_{L}^{{\rm rad}}.
\label{eq:Irad-Jrad-def}
\end{align}
Bearing in mind the above equations jointly with Eqs. \eqref{eq:radiative-source_1PN}, \eqref{eq:I_L and J_L-torsion}, and \eqref{eq:sigma-sigmai-sigmaii-Weyseenhoff},  the 1PN fine-grained form of  $I_{ij}^{{\rm rad}}$,  and the leading-order expressions of $I_{ijk}^{{\rm rad}}$, $I_{ijkl}^{{\rm rad}}$,  $J_{ij}^{{\rm rad}}$, and $J_{ijk}^{{\rm rad}}$ are given by, respectively,
\begin{align}
    I_{ij}^{{\rm rad}}(t)&= \int {\rm d}^3\boldsymbol{x} \Biggl[ x^{\langle ij\rangle } \left(\rho^{\star \star} + \rho_{\rm v}\right)
    \nonumber \\
    & + \dfrac{4}{c^2} \left(s_{il}x^j+s_{jl}x^i-\dfrac{2}{3}\delta^{ij}s_{kl}x^k \right)v^l \Biggr]
    \nonumber \\
    &+ \dfrac{1}{14c^2} \dfrac{{\rm d}^2}{{\rm d}t^2}  \int {\rm d}^3\boldsymbol{x} \, x^{\langle ij\rangle } \boldsymbol{x}^2 \rho^\star 
    \nonumber \\
    &-  \dfrac{20}{21c^2} \dfrac{{\rm d}}{{\rm d}t}  \int {\rm d}^3\boldsymbol{x} \Biggl[ x^{\langle ijk\rangle } v^k \rho^\star 
    \nonumber \\
    &+\dfrac{7}{5} \left(s_{il}x^j+s_{jl}x^i\right)x^l \Biggr] +{\rm O}(c^{-3}),
    \label{eq:I-ij-rad-expression-1}
\\
I_{ijk}^{{\rm rad}}(t)&= \int {\rm d}^3\boldsymbol{x}\, x^{\langle ijk\rangle }\rho^\star +{\rm O}(c^{-2}),
\\
I_{ijkl}^{{\rm rad}}(t)&= \int {\rm d}^3\boldsymbol{x}\, x^{\langle ijkl\rangle }\rho^\star +{\rm O}(c^{-2}),
\\
J_{ij}^{{\rm rad}}(t)&= \int {\rm d}^3\boldsymbol{x}\, \epsilon^{kl\langle i} x^{j\rangle k} \rho^\star v^l
\nonumber \\
& \dfrac{1}{2} \int {\rm d}^3\boldsymbol{x} \left[ \left(2\epsilon^{kl(i}s_{j)l}\right)x^k + \left(2\epsilon^{kl(i} x^{j)}\right)s_{kl}\right] 
\nonumber \\
&+{\rm O}(c^{-2}),
\label{eq:J-ij-rad-expression-1}
\\
J_{ijk}^{{\rm rad}}(t)&= \int {\rm d}^3\boldsymbol{x}\Bigl[\rho^\star x^{\langle i} x^j \epsilon^{k \rangle lp} x^l v^p+ 2 s_{np} \delta_n^{\langle i} x^j \epsilon^{k \rangle lp} x^l
\nonumber \\
&+ s_{lp} x^{\langle i} x^j \epsilon^{k \rangle lp}  \Bigr]  +{\rm O}(c^{-2}),
\label{eq:J-ijk-rad-expression-1}
\end{align}
where we have employed the Gauss theorem to discard the integrals
containing a total divergence (recall that, in our model, both   the spin angular momentum tensor  \eqref{eq:spin-tensor-fluid} and the metric energy-momentum tensor \eqref{eq:T_alpha_beta_fluid} have compact support) and, in Eq. \eqref{eq:J-ij-rad-expression-1}, the identity
\begin{align}
\epsilon^{kl\langle i} \widecheck{x}^{j\rangle k}& = \epsilon^{kl\langle i} x^{j\rangle k}=\epsilon^{kl (i} x^{j) k}.
\end{align}

\subsubsection{The ADM mass and the center of mass}
\label{sec:PP-limit-preliminary}

In Ref. \cite{Paper1}, we have demonstrated that the mass monopole moment  $I$, which can be read off   Eq. \eqref{eq:I_L_torsion} with $l=0$, gives rise to a generalized notion of \emph{ ADM mass} (or \emph{total mass-energy}) of the fluid system and can be written as
\begin{align}
 I\left(t\right) &\equiv M_{\rm ADM}\left(t\right)= \int {\rm d}^3 \boldsymbol{x} \left[ \sigma + \dfrac{1}{c^2}\left(\dfrac{1}{2} \sigma U -\sigma_{ii}\right)\right] 
 \nonumber \\
 &+ {\rm O}(c^{-4}), \label{eq:I-expression-torsion}
\end{align}
and, in addition, satisfies
\begin{align} \label{eq:deriv-t-of-M-ADM}
    \dfrac{{\rm d}}{{\rm d}t} M_{\rm ADM}=0.
\end{align}
In the case of the Weyssenhoff fluid, we find, with the help of Eq. \eqref{eq:sigma-sigmai-sigmaii-Weyseenhoff},
\begin{align} \label{eq:ADM-mass-fluid_1}
    M_{\rm ADM} = \int {\rm d}^3 \boldsymbol{x} \, \rho^{\star \star}  + {\rm O}(c^{-4}),
\end{align}
where we have exploited  Gauss theorem to discard total derivatives  occurring in $\sigma$ and $\sigma_{ii}$ (cf. Eqs. \eqref{Eq:sigma-EC-theory} and \eqref{Eq:sigma-EC-theory-ii}). For this reason, Eq. \eqref{eq:ADM-mass-fluid_1} resembles formally the GR expression \cite{Poisson-Will2014}. The ADM mass \eqref{eq:ADM-mass-fluid_1} assumes the equivalent form
\begin{equation} \label{eq:ADM-mass_2}
    M_{\rm ADM} \equiv m^{\rm mat} + \dfrac{\mathscr{E}}{c^2},
\end{equation}
where
\begin{equation} 
    \mathscr{E} = \mathscr{T} + \mathscr{E}_{\rm p} + \mathscr{E}_{\rm int} + {\rm O}(c^{-2}),
\end{equation}
and
\begin{subequations}
\begin{align}
m^{\rm mat} & \equiv \int {\rm d}^3 \boldsymbol{x} \, \rho^{\star},
\\
\mathscr{T} & \equiv \dfrac{1}{2} \int {\rm d}^3 \boldsymbol{x} \, \rho^{\star} v^2,
\\
\mathscr{E}_{\rm p}  &\equiv -\dfrac{1}{2} \int {\rm d}^3 \boldsymbol{x} \, \rho^{\star} U,
\\
\mathscr{E}_{\rm int}  &\equiv \int {\rm d}^3 \boldsymbol{x} \, \rho^{\star} \Pi,
\label{eq:internal_energy_fluid_integral}
\end{align}
\end{subequations}
denote the total material mass, the translational kinetic energy, the total  gravitational potential energy, and the total internal energy of the fluid system, respectively. It should be noted that 
the contribution of  the spin kinetic energy is included in Eq. \eqref{eq:internal_energy_fluid_integral}.
Furthermore,   Eqs. (\ref{eq:cons-mass-with-rho-star}),  (\ref{eq:0PN-eq2}), and
\begin{align}
    \dfrac{{\rm d}\Pi}{{\rm d}t}= \dfrac{P}{\rho^{\star 2}}\dfrac{{\rm d}\rho^\star}{{\rm d}t}+ {\rm O}(c^{-2}),
\end{align}
which can be obtained through Eqs. (\ref{eq:thermodynamic-relation-3}) and (\ref{eq:rho-star-in-terms-of-rho}), assure the constancy of the ADM mass \eqref{eq:ADM-mass_2}, in agreement with Eq. \eqref{eq:deriv-t-of-M-ADM}.

The case $l=1$ of Eq. \eqref{eq:I_L_torsion} yields the definition of the \emph{dipole moment}, which for the Weyssenhoff fluid becomes
\begin{align}
    I_i(t)= \int {\rm d}^3 \boldsymbol{x} \,x^i \rho^{\star \star} + \dfrac{2}{c^2} \int {\rm d}^3 \boldsymbol{x}\, s_{il}v^l + {\rm O}(c^{-4}),
\end{align}
where we have exploited the Gauss theorem.  

The \emph{position of the center of mass} of the fluid system, which  is defined as $\mathscr{R}^i  \equiv I_i/M_{\rm ADM}$ \cite{Poisson-Will2014,Blanchet2014},  reads as
\begin{align} \label{eq:center-of-mass_1}
\mathscr{R}^i  &= \dfrac{1}{M_{\rm ADM}}\left[\int {\rm d}^3 \boldsymbol{x} \,x^i \rho^{\star \star} + \dfrac{2}{c^2} \int {\rm d}^3 \boldsymbol{x}\, s_{il}v^l \right]
\nonumber \\
&+ {\rm O}(c^{-4}),
\end{align}
and satisfies the relation 
\begin{equation}
    M_{\rm ADM} \dfrac{{\rm d}\mathscr{R}^i}{{\rm d}t}= \mathscr{P}^i + {\rm O}(c^{-4}),
\end{equation}
$\mathscr{P}^i$ being the \emph{conserved ADM three-momentum} of the system (see Ref. \cite{Paper1} for further details).

It should be noted that an explicit spin correction term occurs in Eq. \eqref{eq:center-of-mass_1}. This is similar to the contribution appearing   in the analysis of spinning bodies    within GR  \cite{Poisson-Will2014,Costa2014a}. However, we recall the basic difference between EC model and Einstein theory, as in the former the spin refers to the intrinsic \emph{quantum} angular momentum of elementary particles, while in the latter the spin is related to a macroscopic rotation \cite{Hehl1976_fundations}.

\subsubsection{The coarse-grained description}
\label{sec:coarse-grained-descr}

We are now ready to apply the results of the previous sections to the framework where the 
fine-grained description of the Weyssenhoff fluid is replaced by the  coarse-grained picture involving separated fluid elements,  which are characterized  by a small number of  variables \cite{Poisson-Will2014,Blanchet-Schafer1989}. Therefore,  we consider a setup where the fluid distribution breaks up into a collection of $N$ weakly self-gravitating, slowly moving, widely separated, and isolated spinning components  surrounded by vacuum regions of space. Each component, referred to as body,  is assigned  a label $A=1,2,\dots,N$, and, in our hypotheses, the ratio $\alpha \equiv \lambda/\delta \ll 1$ ($\lambda$ and $\delta$ being the typical size   of $A$  and the typical separation  between the bodies, respectively). The coordinate fluid density can be expressed as $\rho^\star = \sum_{A} \rho^\star_A$, where the sum extends over each body, and $\rho^\star_A$ vanishes everywhere except within the volume  occupied by the body $A$. Furthermore, the (conserved) material mass of  $A$ is $m^{\rm mat}_A = \int_A {\rm d}^3 \boldsymbol{x} \, \rho^\star$, the domain of integration being a time-independent portion of the three-dimensional space which extends slightly beyond the volume occupied by the body $A$ and does not include nor intersect another body within the system (recall that $\rho^\star=\rho^\star_A$ within such a domain).

Similarly, the spin density tensor $s_{ij}$ is zero in the vacuum exterior of the bodies, whereas $s_{ij}=s^A_{ij}$ inside the volume of the body $A$. The spin vector $\boldsymbol{s}_A$ of $A$ is 
defined by
\begin{align} \label{eq:spin-vector-body-A}
    \epsilon_{jki}s_A^i(t)=\int_A {\rm d}^3 \boldsymbol{x} \, s^A_{jk},
\end{align}
and, owing to Eq. \eqref{eq:0PN-eq3}, is conserved modulo $\rm{O}\left(c^{-2}\right)$ corrections, i.e., 
\begin{align} \label{eq:spin-conserved-expr}
    \dfrac{{\rm d}}{{\rm d}t}s^i_A = \rm{O}\left(c^{-2}\right).
\end{align}

Driven by Eqs.  \eqref{eq:center-of-mass_1} and \eqref{eq:ADM-mass-fluid_1}, the center of mass worldline $r^i_A\left(t\right)$ of the body $A$ is defined as
\begin{align}
    r^i_A\left(t\right) &= \dfrac{1}{m_A} \int_A {\rm d}^3 \boldsymbol{x} \,x^i \rho_A^\star \left[1+\dfrac{1}{c^2} \left(\dfrac{w_A^2}{2} + \Pi_A -\dfrac{u_A}{2}\right)\right] 
\nonumber \\    
&+ \dfrac{2}{m_A c^2} \int_A {\rm d}^3 \boldsymbol{x} \, s^A_{il}w^l_A +{\rm O}(c^{-4}),
\label{eq:center-of-mass-bodyA}
\end{align}
where 
\begin{align}
m_A  &=\int_A {\rm d}^3 \boldsymbol{x} \, \rho_A^\star \left[1+\dfrac{1}{c^2} \left(\dfrac{w_A^2}{2} + \Pi_A -\dfrac{u_A}{2}\right)\right] 
\nonumber \\    
&+ {\rm O}(c^{-4}),
\label{eq:total-mass-bodyA}
\end{align}
is the total mass-energy of the body, which, along the same lines as for $M_{\rm ADM}$, can be shown to be conserved. In Eqs. \eqref{eq:center-of-mass-bodyA} and \eqref{eq:total-mass-bodyA},   $\Pi_A$ denotes the specific internal energy of $A$;  furthermore, $u_A$ is the internal (self-gravity) potential, which is readily obtained from the Newtonian potential \eqref{eq:potential-U-Weyssenhoff-fluid}; finally, upon introducing the vector $y^i_A \equiv x^i -r^i_A\left(t\right)$, measuring the position of a fluid element relative to the center of mass $r^i_A\left(t\right)$,  $w^i_A$  is given by $w^i_A \equiv \tfrac{{\rm d}}{{\rm d}t} y^i_A = v^i - v^i_A\left(t\right)$, and represents the velocity of this fluid element relative to the  body velocity $v^i_A\left(t\right) \equiv \tfrac{{\rm d}}{{\rm d}t} r^i_A \left(t\right)$. The above definitions permit renormalizing the internal selfgravity of $A$ into the mass  $m_A$.

\subsubsection{The point-particle limit of the radiative moment}
\label{sec:PP-limit-results}

At this stage, we have all the ingredients needed to perform the point-particle limit of the radiative moments \eqref{eq:I-ij-rad-expression-1}--\eqref{eq:J-ijk-rad-expression-1}. For this reason we employ the same techniques as  in Ref. \cite{Blanchet-Schafer1989}.

In EC theory,  the following crucial identities hold:
\begin{subequations}
\label{eq:p-p-limit-idenities}
\begin{align}
&\int_A {\rm d}^3 \boldsymbol{y}_A y^i_A \rho^\star_A \left[ 1+\dfrac{1}{c^2}\left(\dfrac{w_A^2}{2}+\Pi_A -\dfrac{u_A}{2}\right)\right] 
\nonumber \\
&+ \dfrac{2}{c^2} \int_A {\rm d}^3 \boldsymbol{y}_A s^A_{il}w^l_A = {\rm O}\left(c^{-4}\right),
\label{eq:p-p-limit-idenity1}
\\
& \int_A {\rm d}^3 \boldsymbol{y}_A y^i_A \rho^\star_A = {\rm O}\left(c^{-2}\right),
\label{eq:p-p-limit-idenity2}
\\
& \int_A {\rm d}^3 \boldsymbol{y}_A w^i_A \rho^\star_A = {\rm O}\left(c^{-2}\right),
\label{eq:p-p-limit-idenity3}
\\
&\dfrac{1}{2} \dfrac{{\rm d}^2}{{\rm d}t^2} \int_A {\rm d}^3 \boldsymbol{y}_A \boldsymbol{y}^2_A \rho^\star_A =  \int_A {\rm d}^3 \boldsymbol{y}_A  \rho^\star_A  \left(w_A^2 -\dfrac{u_A}{2}+\dfrac{3P_A}{\rho^\star_A}\right)
\nonumber \\
&+ {\rm O}\left(c^{-2}\right),
\label{eq:p-p-limit-idenity4}
\end{align}
\end{subequations}
where $P_A$ is the pressure within body $A$ and all functions are supposed to depend on  $t$ and  the position $\boldsymbol{y}_A + \boldsymbol{r}_A(t)$.
Equation \eqref{eq:p-p-limit-idenity1} is a consequence of Eqs. \eqref{eq:center-of-mass-bodyA} and \eqref{eq:total-mass-bodyA};  Eq. \eqref{eq:p-p-limit-idenity2} stems from Eq. \eqref{eq:p-p-limit-idenity1} and, in turn, Eq. \eqref{eq:p-p-limit-idenity3} can be obtained by evaluating the time derivative of Eq. \eqref{eq:p-p-limit-idenity2}; finally, Eq. \eqref{eq:p-p-limit-idenity4} derives from the virial theorem, which reads as (cf. Eq. \eqref{eq:sigma-sigmai-sigmaii-Weyseenhoff})
\begin{align}
    \dfrac{1}{2} \dfrac{{\rm d}^2}{{\rm d}t^2} \int {\rm d}^3 \boldsymbol{x}\, \boldsymbol{x}^2 \sigma = \int {\rm d}^3 \boldsymbol{x} \left( \sigma_{jj} -\dfrac{1}{2}\sigma U \right)+  {\rm O}\left(c^{-2}\right).
\end{align}

By  supposing that all bodies are \emph{spherically symmetric} and in  \emph{static equilibrium},  we find that the point-particle counterpart  of the radiative  mass quadrupole moment \eqref{eq:I-ij-rad-expression-1}  reads as 
\begin{align} 
I^{\rm rad}_{ij}&=\sum_A m_A \Biggl\{ r_A^{\langle i}r_A^{j\rangle }\left[1+\frac{1}{c^2}\left(\frac{3}{2} v_A^2 \right. \right.
\nonumber \\
&\left.\left. -\sum_{B\neq A}\frac{G m_B}{|\boldsymbol{r}_A-\boldsymbol{r}_B|}\right)\right]+\frac{1}{14c^2}\frac{\rm{d}^2}{\rm{d}t^2}\left(r_A^2 r_A^{\langle i}r_A^{j\rangle }\right) 
\nonumber \\
&-\frac{20}{21c^2}\frac{\rm{d}}{\rm{d}t}\left(v_A^kr_A^{\langle i}r_A^jr_A^{k\rangle }\right)  \Biggr\} 
\nonumber \\
&+ \sum_{A} \Biggl\{ \dfrac{4}{c^2}  \Bigl[ \left(\boldsymbol{v}_A \times \boldsymbol{s}_A\right)^i r^j_A+\left(\boldsymbol{v}_A \times \boldsymbol{s}_A\right)^jr^i_A 
\nonumber \\
&  -\dfrac{2}{3}\delta^{ij} \left(\boldsymbol{v}_A \times \boldsymbol{s}_A\right) \cdot \boldsymbol{r}_A\Bigr] -  \dfrac{4}{3c^2}  \dfrac{{\rm d}}{{\rm d}t} \Bigl[\left(\boldsymbol{r}_A \times \boldsymbol{s}_A\right)^i r^j_A 
\nonumber \\
&+\left(\boldsymbol{r}_A \times \boldsymbol{s}_A\right)^j r^i_A\Bigr]\Biggr\} + \rm{O}\left(c^{-3}\right);
\label{eq:I-ij-rad-point-particle-limit}
\end{align}
moreover, for the the mass octupole and  current quadrupole, the point-particle procedure yields, respectively, 
\begin{align}
I^{\rm rad}_{ijk}&=\sum_A m_A r_A^{\langle i}r_A^jr_A^{k\rangle }+\rm{O}\left(c^{-2}\right),
\label{eq:I-ijk-rad-point-particle-limit}
\\
J^{\rm rad}_{ij} &=\sum_A m_A \epsilon^{kl\langle i}r_A^{j\rangle }r_A^kv_A^l+
\nonumber \\
&\dfrac{1}{2}\sum_A \left[3 \left(s^i_A r^j_A+s^j_A r^i_A\right)-2 \delta^{ij} \boldsymbol{s}_A \cdot \boldsymbol{r}_A\right] +\rm{O}\left(c^{-2}\right);
\label{eq:J-ij-rad-point-particle-limit}
\end{align}
the mass $2^4$-pole and the current octupole give, respectively,
\begin{align}
   I^{\rm rad}_{ijkl}&=\sum_A m_A r_A^{\langle i}r_A^jr^k_Ar_A^{l\rangle }+\rm{O}\left(c^{-2}\right),
\label{eq:I-ijkl-rad-point-particle-limit} 
\\
J^{\rm rad}_{ijk} &=\sum_A \Bigl[ m_A r_A^{\langle i} r^j_A \epsilon^{k \rangle lp} r_A^l v^p_A+ 2 \Bigl(r^n_A s^q_A \, \delta_n^{\langle i}r^j_A \delta^{k \rangle}_q
\nonumber \\
&-\boldsymbol{r}_A \cdot \boldsymbol{s}_A \, \delta_n^{\langle i}r^j_A \delta^{k \rangle}_n + s_A^q \, r_A^{\langle i} r^j_A \delta_q^{k \rangle}\Bigr) \Bigr] +\rm{O}\left(c^{-2}\right).
\label{eq:J-ijk-rad-point-particle-limit}
\end{align}

Equations \eqref{eq:I-ij-rad-point-particle-limit}--\eqref{eq:J-ij-rad-point-particle-limit} are required for the computation of the instantaneous luminosity \eqref{eq:power-radiated-1PN}, whereas the whole set of radiative moments \eqref{eq:I-ij-rad-point-particle-limit}--\eqref{eq:J-ijk-rad-point-particle-limit} appears in the asymptotic waveform \eqref{eq:gravitational_wave_amplitude} (cf. Eq. \eqref{eq:Irad-Jrad-def}). It is  clear that  EC theory brings in  explicit corrections due to the spin $\boldsymbol{s}_A$ of the body $A$  in $I^{\rm rad}_{ij}$,  $J^{\rm rad}_{ij}$, and $J^{\rm rad}_{ijk}$.

We recall that, in deriving  Eqs.  \eqref{eq:I-ij-rad-point-particle-limit}--\eqref{eq:J-ijk-rad-point-particle-limit}, we have neglected terms $\rm{O}\left(\alpha^{2}\right)$. 

\subsection{Binary systems}
\label{Sec:Binary-System}

In this section, we set out the features of the special case of binary systems, i.e., a collection of $N=2$ weakly self-gravitating, slowly moving, and widely separated spinning fluid bodies having masses $m_1$, $m_2$ (with $m_1\ge m_2$), position vectors $\boldsymbol{r}_1$,  $\boldsymbol{r}_2$, velocities $\boldsymbol{v}_1$, $\boldsymbol{v}_2$, and spin vectors $\boldsymbol{s}_1$,  $\boldsymbol{s}_2$. Let    $M\equiv m_1+m_2$, $\mu\equiv \tfrac{m_1m_2}{M}$, and   $\nu\equiv \tfrac{\mu}{M}$ denote the total mass, the reduced mass, and the symmetric mass ratio of the system, respectively.
As in the Newtonian framework \cite{Poisson-Will2014,Damour1985}, the study of the dynamics is simplified once the origin of the coordinate frame is attached  to the barycenter $\mathscr{R}^i$ of the system and the position of each body is determined in terms of their separation vector. Accordingly, 
we introduce the instantaneous relative position vector $\boldsymbol{R}$ and the instantaneous relative velocity vector $\boldsymbol{V}$\footnote{The relative velocity vector has the same notation as   the retarded potential $V$ (cf. Eq. \eqref{eq:definition-of-retarded-potential-V}); however,  the latter will not be considered in the following sections, and hence no confusion arises in using the same symbol for both quantities.} of the two objects 
\begin{subequations}
\label{eq:relative-position-vector-def&relative-velocity-vector-def}
\begin{align} 
\boldsymbol{R}(t)&\equiv \boldsymbol{r}_1(t)-\boldsymbol{r}_2(t),
\label{eq:relative-position-vector-def}
\\
\boldsymbol{V}(t)&\equiv \dfrac{{\rm d}}{{\rm d}t}\boldsymbol{R}(t)=\boldsymbol{v}_1(t)-\boldsymbol{v}_2(t).
\label{eq:relative-velocity-vector-def}
\end{align}
\end{subequations}
Moreover, starting from Eq. \eqref{eq:center-of-mass_1}, we find that, within the coarse-grained description of the fluid and resorting to the same techniques as in the last sections, the position of the barycenter of a binary system is defined by (we discard, like before, $\rm{O}\left(\alpha^{2}\right)$ corrections)
\begin{align} \label{eq:center-of-mass-2-bodies}
M_{\rm ADM}\mathscr{R}^i &=   m_1 \Biggl[ 1+
 \dfrac{1}{2c^2} \Biggl( v_1^2   -  \dfrac{Gm_2}{R}\Biggl)\Biggr]r^i_1 
\nonumber \\
&+ \dfrac{2}{c^2}\left(\boldsymbol{v}_1 \times \boldsymbol{s}_1\right)^i  
   \nonumber \\
   &+  m_2 \Biggl[ 1
+ \dfrac{1}{2c^2} \Biggl( v_2^2   -  \dfrac{Gm_1}{R}\Biggl)\Biggr]r^i_2 
\nonumber \\
&+ \dfrac{2}{c^2}\left(\boldsymbol{v}_2 \times \boldsymbol{s}_2\right)^i  +\rm{O}\left(c^{-4}\right).
\end{align}
By employing a  post-Galilean transformation (i.e., a particular subclass of general PN transformations \cite{Poisson-Will2014}), it is always possible to define the center of mass frame of the system by 
setting $\mathscr{R}^i=0$ and $\mathscr{P}^i=0$. Therefore, by means of  Eq. \eqref{eq:center-of-mass-2-bodies}, we find that in a mass-centered coordinate system the motion of the bodies is related to their relative motion by the following relations: 
\begin{subequations}
\label{eq:position-vectors-r1-r2-with-spin}
\begin{align}
\boldsymbol{r}_1(t)&=\left[\frac{\mu}{m_1}+\frac{\mu (m_1-m_2)}{2M^2c^2}\left(V^2-\frac{GM}{R}\right)\right]\boldsymbol{R}(t)
\notag\\
&+\frac{2 \nu}{c^2}\left[\dfrac{\boldsymbol{s}_1(t)}{m_1} -\dfrac{\boldsymbol{s}_2(t)}{m_2}\right]\times \boldsymbol{V}(t)+{\rm O}\left(c^{-4}\right),
\\
\boldsymbol{r}_2(t)&=\left[-\frac{\mu}{m_2}+\frac{\mu (m_1-m_2)}{2M^2c^2}\left(V^2-\frac{GM}{R}\right)\right]\boldsymbol{R}(t)\notag\\
&+\frac{2 \nu}{c^2}\left[\dfrac{\boldsymbol{s}_1(t)}{m_1} -\dfrac{\boldsymbol{s}_2(t)}{m_2}\right]\times \boldsymbol{V}(t)+{\rm O}\left(c^{-4}\right).
\end{align}
\end{subequations}

If we replace Eq. \eqref{eq:position-vectors-r1-r2-with-spin} in Eqs. \eqref{eq:I-ij-rad-point-particle-limit}--\eqref{eq:J-ijk-rad-point-particle-limit}, we find that the general form of the radiative moments for  binaries of spinning objects reads as 
\begin{align}
I^{\rm rad}_{ij}& =\mu R_{\langle ij\rangle }\left[1+\frac{3}{2c^2}(1-3\nu)V^2-\frac{(1-2\nu)}{c^2}\frac{GM}{R}\right]\notag\\
&-\frac{\mu(1-3\nu)}{21c^2}\left[20\frac{{\rm d}}{{\rm d}t}(V_kR_{\langle ijk\rangle })-\frac{3}{2}\frac{{\rm d}^2}{{\rm d}t^2}(R^2R_{\langle ij\rangle })\right]\notag\\
&+\frac{8\mu^2}{c^2}\left\{\frac{(\boldsymbol{V}\times\boldsymbol{s}_1)^{\langle i}R^{j\rangle }}{m_1^2}+\frac{(\boldsymbol{V}\times\boldsymbol{s}_2)^{\langle i}R^{j\rangle }}{m_2^2}\right.\notag\\
&\left.-\frac{1}{3}\frac{{\rm d}}{{\rm d}t}\left[\frac{(\boldsymbol{R}\times\boldsymbol{s}_1)^{(i}R^{j)}}{m_1^2}+\frac{(\boldsymbol{R}\times\boldsymbol{s}_2)^{(i}R^{j)}}{m_2^2}\right]\right\}\notag\\
&+{\rm O}\left(c^{-3}\right),
\label{I-ij-rad-pp-limit-expression-2}\\
I^{\rm rad}_{ijk}& =-\mu\sqrt{1-4\nu}R_{\langle ijk\rangle }+{\rm O}\left(c^{-2}\right),
\\
J^{\rm rad}_{ij}& =-\mu\sqrt{1-4\nu}\epsilon_{kl\langle i}R_{j\rangle k}V_l\notag\\
&+3\mu\left(\frac{s_1^{\langle i}R^{j\rangle }}{m_1}-\frac{s_2^{\langle i}R^{j\rangle }}{m_2}\right)
+{\rm O}\left(c^{-2}\right),
\label{J-ij-rad-pp-limit-expression-2}
\\
I^{\rm rad}_{ijkl}& =\mu(1-3\nu)R_{\langle ijkl\rangle} +{\rm O}\left(c^{-2}\right),
\\
J^{\rm rad}_{ijk}&=\mu(1-3\nu)R_{\langle ij}\epsilon_{k\rangle lp}R_l V_p\notag\\
&+\sum_{A=1}^{2}\frac{2\mu^2}{m_A^2} \Bigl(R^n s^q_A \, \delta_n^{\langle i}R^j \delta^{k \rangle}_q-\boldsymbol{R} \cdot \boldsymbol{s}_A \, \delta_n^{\langle i}R^j \delta^{k \rangle}_n
\nonumber \\
& \hspace{1.5cm}+ s_A^q \, R^{\langle i} R^j \delta_q^{k \rangle}\Bigr)+{\rm O}\left(c^{-2}\right).
\label{J-ijk-rad-pp-limit-expression-2}
\end{align}

The above equations completely determine the 1PN generation of GWs from binary systems in EC theory. 

\section{First application to binary neutron star systems}
\label{Sec:Application}

The theoretical pattern  developed in the previous sections will be now  applied to the study of binary NSs. A full analysis requires the knowledge of the 1PN dynamics  in EC theory, which, at the moment, is not at our disposal. Despite that, we can provide a first estimation  by  following a \emph{hybrid approach}. We exploit the EC definition of center of mass and the general expression of the EC radiative moments obtained in  Sec. \ref{Sec:Binary-System}, along with 
the conservation equation of the spin vector  (cf. Eqs. \eqref{eq:0PN-eq3} and \eqref{eq:spin-conserved-expr}); furthermore, we consider the \emph{quasi-elliptic 1PN-accurate GR motion of a binary system} determined by Damour and Deruelle \cite{Damour1985} (see Sec. \ref{sec:motion_1PN_GR}).

We note that, in the full EC framework, only the time derivatives of the radiative mass quadrupole moment
 $I_{ij}^{\rm rad}$  will contain new $\rm{O}(c^{-2})$ spin-dependent terms,  whereas the other moments will remain unaffected. Indeed, as pointed out before, the EC 0PN-accurate translational motion   coincides with the Newtonian Euler equation owing to the Frenkel condition (cf. Eq. \eqref{eq:0PN-eq2}).

In this hybrid setup, we obtain the explicit expressions of the flux and the gravitational waveform (see Sec. \ref{sec:flux-gwf}). We  conclude the section with a numerical estimate concerning the EC corrections by examining  binary NS systems   (see Sec. \ref{sec:num_sim}). 

Any effect of GW back-reaction on the source dynamics will be neglected. This hypothesis, which is valid to a good approximation in some astrophysical GW sources (see e.g. Refs. \cite{Noutsos2020,Weisberg2010}), permits to derive a first model for the description of  GW phenomena in EC theory. Hereafter, we no longer mention  ${\rm O}\left(c^{-n}\right)$ terms.

\subsection{The Damour-Deruelle solution}
\label{sec:motion_1PN_GR}

We consider, within the hybrid approach set forth above, a binary system consisting of two PN widely separated spinning bodies having masses $m_1$, $m_2$ and  spin vectors $\boldsymbol{s}_1$, $\boldsymbol{s}_2$. As pointed out before, we assume that their dynamics is governed by the Damour-Deruelle solution, which we  now briefly outline.

The problem of solving the motion of the binary system can be reduced to the simpler equivalent task of determining the relative motion in the PN center of mass frame (which can be defined by means of the results of Sec. \ref{Sec:Binary-System}). For this reason, given a harmonic coordinate system, we  define, in the same way as before, the instantaneous relative position vector $\boldsymbol{R}$ and the instantaneous relative velocity vector $\boldsymbol{V}$ (cf. Eq. \eqref{eq:relative-position-vector-def&relative-velocity-vector-def}). 

The total energy $E$ and the total angular momentum $\boldsymbol{J}$ of the system (which are conserved in the Damour-Deruelle dynamics) read as, respectively,  
\begin{align}
E &=\frac{1}{2}V^2-\frac{GM}{R}+ \dfrac{1}{c^2} \Biggl\{ \frac{3}{8}V^4 (1-3\nu)
\notag\\
&+\frac{GM}{2R}\left[(3+\nu)V^2+\nu\left(\frac{\boldsymbol{R}\cdot\boldsymbol{V}}{R}\right)^2+\frac{GM}{R}\right] \Biggr\},
\label{eq:first-integral-GR-E}
\\
\boldsymbol{J}&=\boldsymbol{R}\times\boldsymbol{V}\left\{ 1+\dfrac{1}{c^2}\left[\frac{(1-3\nu)}{2}V^2+(3+\nu)\frac{GM}{R}\right]\right\}.
\label{eq:first-integral-GR-J}
\end{align}
The motion takes place in the plane orthogonal to $\boldsymbol{J}$, which is henceforth supposed to be directed along the $z$-axis. We can thus introduce  polar coordinates $(R,\varphi)$ and write  the  equation  of the PN \emph{relative orbit} as
\begin{equation}
\label{eq:relative-orbit-R-varphi}
R(\varphi)=\left(a_{\rm R}-\frac{G\mu}{2c^2}\right)\frac{1-e_\varphi^2}{1+e_\varphi \cos\left(\frac{\varphi-\varphi_{\rm in}}{K}\right)}+\frac{G\mu}{2c^2},
\end{equation}
where $\varphi_{\rm in} $ is the initial angle and   the following orbital parameters have been introduced: 
\begin{subequations}
\label{eq:orbital parameters}
\begin{align}
a_R&=-\frac{GM}{2E}\left[1-\frac{1}{2}(\nu-7)\frac{E}{c^2}\right],\\
e_R&=\left\{1+\frac{2E}{G^2M^2}\frac{\left[1+\frac{5}{2}\left(\nu-3\right)\frac{E}{c^2}\right]}{\left[J^2+(\nu-6)\frac{G^2M^2}{c^2}\right]^{-1}}\right\}^{1/2},\\
e_\varphi&=e_R\left(1+\frac{G\mu}{2a_Rc^2}\right),\\
K&=\frac{J}{(J^2-6G^2M^2/c^2)^{1/2}}. \label{eq:K}
\end{align}
\end{subequations}

\subsection{The flux and the gravitational waveform}
\label{sec:flux-gwf}

It follows from   the definition \eqref{eq:Irad-Jrad-def}  that the instantaneous luminosity \eqref{eq:power-radiated-1PN} attains the equivalent form (restoring,  for a while, the ${\rm O}(c^{-n})$ terms)
\begin{align} 
\label{eq:power-radiated-1PN_equivalent-expr}
   \mathcal{F}(t) & = \dfrac{G}{c^5} \Biggl\{\dfrac{1}{5}\overset{(3)}{I}{}_{ij}^{{\rm rad}}\overset{(3)}{I}{}^{{\rm rad}}_{ij} + \dfrac{1}{c^2} \Biggl[ \dfrac{1}{189} \overset{(4)}{I}{}^{{\rm rad}}_{ijk} \overset{(4)}{I}{}^{{\rm rad}}_{ijk}
 \nonumber \\
&+ \dfrac{16}{45} \overset{(3)}{J}{}^{\rm rad}_{ij}\overset{(3)}{J}{}_{ij}^{\rm rad} \Biggr]
+ {\rm O}(c^{-4}) \Biggr \},
\end{align}
whereas the asymptotic amplitude (\ref{eq:gravitational_wave_amplitude}) becomes 
\begin{align} \label{eq:gravitational_wave_amplitude-equival-express}
    \mathscr{H}_{ij}^{\rm TT}(x^\mu) & = \dfrac{2G}{c^4 \vert\boldsymbol{x}\vert} \mathscr{P}_{ijkl}(\boldsymbol{n})  \Biggr\{ \overset{(2)}{I}{}_{kl}^{{\rm rad}}(u) 
    \nonumber \\
    & + \dfrac{1}{c} \left[  \dfrac{1}{3} n_a  \overset{(3)}{I}{}_{kla}^{{\rm rad}}(u) +\dfrac{4}{3} n_b \epsilon_{ab(k} \overset{(2)}{J}{}_{l)a}^{{\rm rad}}(u)  \right]
    \nonumber \\
    & +\dfrac{1}{c^2} \left[\dfrac{1}{12}n_{a}n_{b}   \overset{(4)}{I}{}_{klab}^{{\rm rad}}(u) \right.
    \nonumber \\
     &\left. + \dfrac{1}{2}n_{b}n_{c} \epsilon_{ab(k}   \overset{(3)}{J}{}_{l)ac}^{{\rm rad}}(u)  \right]
      + {\rm O}(c^{-3}) \Biggr\},
\end{align}
where    $u\equiv t-\vert\boldsymbol{x}\vert/c$,  $\boldsymbol{n}\equiv \boldsymbol{x}/\vert\boldsymbol{x}\vert$, and the radiative  moments  have been derived in Eqs. \eqref{I-ij-rad-pp-limit-expression-2}--\eqref{J-ijk-rad-pp-limit-expression-2}. We also note that in Eqs. \eqref{eq:power-radiated-1PN_equivalent-expr} and \eqref{eq:gravitational_wave_amplitude-equival-express} we have exploited the fact that, at this order, there is no difference between the harmonic and the radiative coordinates \cite{Blanchet-Damour1989,Blanchet2014}.

At this stage, we can compute $  \mathcal{F}$ and $\mathscr{H}_{11}^{\rm TT}$ by exploiting the Damour-Deruelle solution. 

The \emph{instantaneous luminosity} \eqref{eq:power-radiated-1PN_equivalent-expr} can be written as
\begin{align} \label{eq:luminosity-plot}
\mathcal{F}(t)=\mathcal{F}_{\rm GR}(t)+\mathcal{F}_{\rm EC}(t),
\end{align}
where the expression of the  GR flux can be found in Ref. \cite{Blanchet-Schafer1989}, while the EC contribution can be obtained after a lengthy calculation by exploiting the results of Secs. \ref{Sec:Binary-System} and \ref{sec:motion_1PN_GR} . In the hypotheses that the motion occurs in the $xy$-plane  (i.e., $R_z=V_z=0$) and  the spins of the two bodies  are aligned with the total angular momentum  $\boldsymbol{J}$  (i.e., $s_{x1}=s_{x2}=s_{y1}=s_{y2}=0$),  we find for the EC luminosity
\begin{align}
\mathcal{F}_{\rm EC}(t)&=
\frac{8 G^3}{15 c^7 R^8} \biggr{\{}8 G \mu M R \left(m_1^2 s_{\rm z2}+m_2^2 s_{\rm z1}\right)\notag\\
&\times(R_x V_y-R_y V_x)+\frac{m_1^3 s_{\rm z2}}{M} \biggr{[}m_2 R_x^3 \left(4 V_x^2 V_y\right.\notag\\
&\left.+58 V_y^3\right)+R_x^2 \left(3 s_{\rm z2} \left(4 V_x^2+V_y^2\right)\right.\notag\\
&\left.-2 m_2 R_y V_x \left(2 V_x^2+83 V_y^2\right)\right)\notag\\
&+2 R_x R_y V_y \left(m_2 R_y \left(83 V_x^2+2 V_y^2\right)+9 s_{\rm z2} V_x\right)\notag\\
&+R_y^2 \left(3 s_{\rm z2} \left(V_x^2+4 V_y^2\right)-2 m_2 R_y V_x \left(29 V_x^2\right.\right.\notag\\
&\left.\left.+2 V_y^2\right)\right)\biggr{]}+\frac{3 m_2^3 s_{\rm z1}^2 }{M}\biggr{[}R_x^2 \left(4 V_x^2+V_y^2\right)\notag\\
&+6 R_x R_y V_x V_y+R_y^2 \left(V_x^2+4 V_y^2\right)\biggr{]}\notag\\
&+\mu  m_1 \biggr{[}R_x^2 \left(4 V_x^2+V_y^2\right)+6 R_x R_y V_x V_y\notag\\
&+R_y^2 \left(V_x^2+4 V_y^2\right)\biggr{]} \left(-2 m_2 R_x V_y (s_{\rm z1}+s_{\rm z2})\right.\notag\\
&\left.+2 m_2 R_y s_{\rm z1} V_x+2 m_2 R_y s_{\rm z2} V_x-6 s_{\rm z1} s_{\rm z2}+3 s_{\rm z2}^2\right)\notag\\
&+\mu  m_2 s_{\rm z1} \biggr{[}V_x^2 \left(3 \left(4 R_x^2+R_y^2\right) (s_{\rm z1}-2 s_{\rm z2})\right.\notag\\
&\left.-2 m_2 R_y V_x \left(2 R_x^2+29 R_y^2\right)\right)+V_y^2 \left(3 \left(R_x^2+4 R_y^2\right)\right.\notag\\
&\left.\times(s_{\rm z1}-2 s_{\rm z2})-2 m_2 R_y V_x \left(83 R_x^2+2 R_y^2\right)\right)\notag\\
&+2 R_x V_x V_y \left(2 m_2 R_x^2 V_x+83 m_2 R_y^2 V_x\right.\notag\\
&\left.+9 R_y (s_{\rm z1}-2 s_{\rm z2})\right)\notag\\
&+2 m_2 R_x V_y^3 \left(29 R_x^2+2 R_y^2\right)\biggr{]}\biggr{\}}.
\label{eq:flux-EC-in-plot}
\end{align}

The general form of the gravitational waveform can be obtained after a long calculation by considering the time derivatives of the radiative moments appearing in Eq. \eqref{eq:gravitational_wave_amplitude-equival-express}.  In this section,  we give  the  component $\mathscr{H}_{11}^{\rm TT}$ of the asymptotic amplitude in the hypotheses that the GW propagates along the direction $\boldsymbol{n}=(0,0,1)$ and, like before,   the spins of the two bodies are aligned and orthogonal to the $xy$-plane of  motion.  The resulting expression  can be written as the sum of the GR and the EC contributions, i.e.,  (in order to ease the notation, henceforth we write $\mathscr{H}_{11} \equiv \mathscr{H}_{11}^{\rm TT}$)
\begin{align} \label{eq:GWF-plot}
\mathscr{H}_{11}(t)&=\mathscr{H}_{11}^{\rm GR}(t)+\mathscr{H}_{11}^{\rm EC}(t),    
\end{align}
where 
\begin{align} 
\mathscr{H}_{11}^{\rm GR}(t)&=
\frac{G \mu}{3d_{\rm so}R^6c^6} \Biggl\{29G^2 M^2 \left(R_x^4-R_y^4\right) +3R^6  
\nonumber \\
& \times \left(V_x^2-V_y^2\right)  \left[2c^2 + V^2 \left(\dfrac{\mu^2}{m_1^2}+\dfrac{\mu^2}{m_2^2}-\nu\right)\right]
\nonumber \\
&-3G M R \Biggl[2c^2 \left(R_x^4 -R_y^4\right)+\left(\dfrac{\mu^2}{m_2^2} + \dfrac{\mu^2}{m_1^2}\right) 
\nonumber \\
& \times \Bigl(2R_x R_y V_x V_y \left(R_y^2-R_x^2\right) +3R_x^2 R_y^2 \left(V_y^2-V_x^2\right)
\nonumber \\
&+ R_x^4 \left(2V_y^2-3V_x^2\right)+R_y^4 \left(3V_y^2-2V_x^2\right) \Bigr)
\nonumber \\
+&\nu \Bigl( 2R_x R_y V_x V_y \left(R_x^2-R_y^2\right) -17R_x^2 R_y^2 \left(V_x^2-V_y^2\right)
\nonumber \\
&+ R_x^4 \left(7V_y^2-8V_x^2\right)+R_y^4 \left(8V_y^2-7V_x^2\right)\Bigr)
\Biggr]\Biggr\},
\label{eq:H-11-GR-in-plot}
\\
\mathscr{H}_{11}^{\rm EC}(t)&=-\dfrac{8 \mu G^2}{c^6d_{\rm so}R^5} \left(\dfrac{m_2}{m_1}s_{\rm z1}+\dfrac{m_1}{m_2}s_{\rm z2}\right)\biggl[R_x^2 \left(2R_xV_y \right.
\nonumber \\
& \left. -R_yV_x \right)+R_y^2 \left(2R_yV_x-R_xV_y\right)\biggr],
\label{eq:H-11-EC-in-plot}
\end{align}
$d_{\rm so}$ being the (constant) distance to the astrophysical source.

\subsection{Numerical estimates}
\label{sec:num_sim}

In this section, we deal with  binary NS  systems and 
provide some numerical estimates of the contributions introduced by EC model in the flux and the  waveform.  In Sec. \ref{sec:par_setting}, we set the parameters which are necessary to perform the numerical computations. These  are then  discussed  in Sec. \ref{sec:NS}. Hereafter, a dot signifies a differentiation with respect to the $t$ variable.

\subsubsection{Parameter setting}
\label{sec:par_setting}

The set of \emph{initial conditions} characterizing our numerical investigation  is represented by $(R_{\rm in},\varphi_{\rm in},\dot{R}_{\rm in},\dot{\varphi}_{\rm in})$. The initial radius $R_{\rm in}$ is expressed in terms of $R_{\rm g}$, with $R_{\rm g}\equiv GM/c^2$; for the initial angle and radial  velocity,  we  assume  $\varphi_{\rm in}=0$ and   $\dot{R}_{\rm in}=0$, respectively;  $\dot{\varphi}_{\rm in}$ is a fraction of the Keplerian velocity, i.e.,  
\begin{equation}\label{eq:dot-varphi-in}
\dot{\varphi}_{\rm in}=\beta \sqrt{\frac{GM}{R_{\rm in}^3}},  \qquad 0<\beta\le1,   
\end{equation}
where for $\beta=1$ the Newtonian eccentricity $e_0$ vanishes (leading to circular orbits), whereas in the limiting case  $\beta\to0$ we have $e_0\to1$. Therefore, given these premises, the initial conditions are specified once we assign $M$, $R_{\rm in}$, and $\beta$. 

A crucial point of our analysis regards the spins of the NSs. These are modeled as follows
\begin{align}
\label{eq:szi-components}
s_{\rm zi}=  n\hbar\frac{4\pi}{3}\left(\frac{6Gm_i}{c^2}\right)^3, \quad i=1,2,
\end{align}
where, following Ref. \cite{Paper1}, $n=10^{44}\ \mbox{m}^{-3}$ is estimated as the inverse of the nucleon volume. Therefore, if the masses $m_1$ and $m_2$ are known, then the spin components $s_{\rm z1} $ and $s_{\rm z2}$ can be  immediately calculated. 

In order to gain useful information about the binary system's dynamics, we determine  the minimum, average, and maximum values  of the relative radius (i.e., $R_{\rm min},R_{\rm av},R_{\rm max}$). Furthermore,  to perform some consistency checks, we define  a set of  parameters, which must be less than 1  due to the hypotheses underlying our model; first of all, the  slow-motion condition demands that we compute the maximum values $v_1^{\rm max}/c,v_2^{\rm max}/c$ attained by the ratios $v_1/c,v_2/c$, respectively (the velocities $v_1$ and $v_2$ of the two bodies can be obtained starting from Eq. \eqref{eq:position-vectors-r1-r2-with-spin}); to verify whether the two bodies  remain widely separated, we calculate 
\begin{equation}
\alpha_i=\frac{12Gm_i/c^2}{R_{\rm min}},\qquad i=1,2; 
\end{equation}
finally, we monitor the strength of the gravitational field  through the factor
\begin{equation}
\gamma=\frac{GM}{c^2R_{\rm min}}.    
\end{equation}

The  values of the  aforementioned variables, along with other quantities characterizing the binary NS   system to be  investigated in Sec. \ref{sec:NS},  are listed in Table \ref{tab:table1}.

\subsubsection{Discussion of the results}
\label{sec:NS}

We consider a gravitational system consisting of two NSs, whose parameters can be found in  Table \ref{tab:table1}.

\setlength{\doublerulesep}{2\arrayrulewidth}
\begin{table}[h!]
    \centering
    \begin{tabular}{|c|c|c|}
         \hline
         & &\\
        {\bf PARAMETERS} & {\bf UNITS} &\hspace{0.3cm} {\bf VALUES }\hspace{0.3cm}\\
        & &\\
         \hline \hline
        $m_1 $ & $M_\odot$ & 1.60  \\
        $m_2$ & $M_\odot$ & 1.17\\
        $M$ & $M_\odot$ & 2.77 \\
        $s_{\rm z1}$ & $\hbar$ &$1.21\times10^{57}$\\
        $s_{\rm z2}$ & $\hbar$ &$4.73\times10^{56}$\\
        $d_{\rm so}$& ${\rm Mpc}$ & 40.00\\
        $R_{\rm g}$ & m & $4.11\times10^3$\\  
        $R_{\rm in}$ & $R_{\rm g}$ & $2.00\times10^5$\\
        $\beta$ & & 0.70 \\
        $e_0$ & & 0.51 \\
        \hline
        $R_{\rm min}$ & $R_{\rm g}$ & $0.65\times10^5$ \\
        $R_{\rm av}$ & $R_{\rm g}$ & $1.14\times10^5$ \\      
        $R_{\rm max}$ & $R_{\rm g}$ & $2.00\times10^5$ \\
        \hline
        $v_1^{\rm max}$ & $c$ & $2.03 \times 10^{-3}$ \\
        $v_2^{\rm max}$ & $c$ & $2.79 \times 10^{-3}$ \\  
        $\alpha_1$ & & $1.07 \times 10^{-4}$ \\
        $\alpha_2$ & & $7.81 \times 10^{-5}$ \\        
        $\gamma$ & & $1.54 \times 10^{-5}$ \\
        \hline
    \end{tabular}
    \caption{List of parameters of the binary NS  system analyzed in Sec. \ref{sec:NS}.}
    \label{tab:table1}
\end{table}
\begin{figure}[th!]
\centering
\hbox{\hspace{-0.24cm}\includegraphics[scale=0.41]{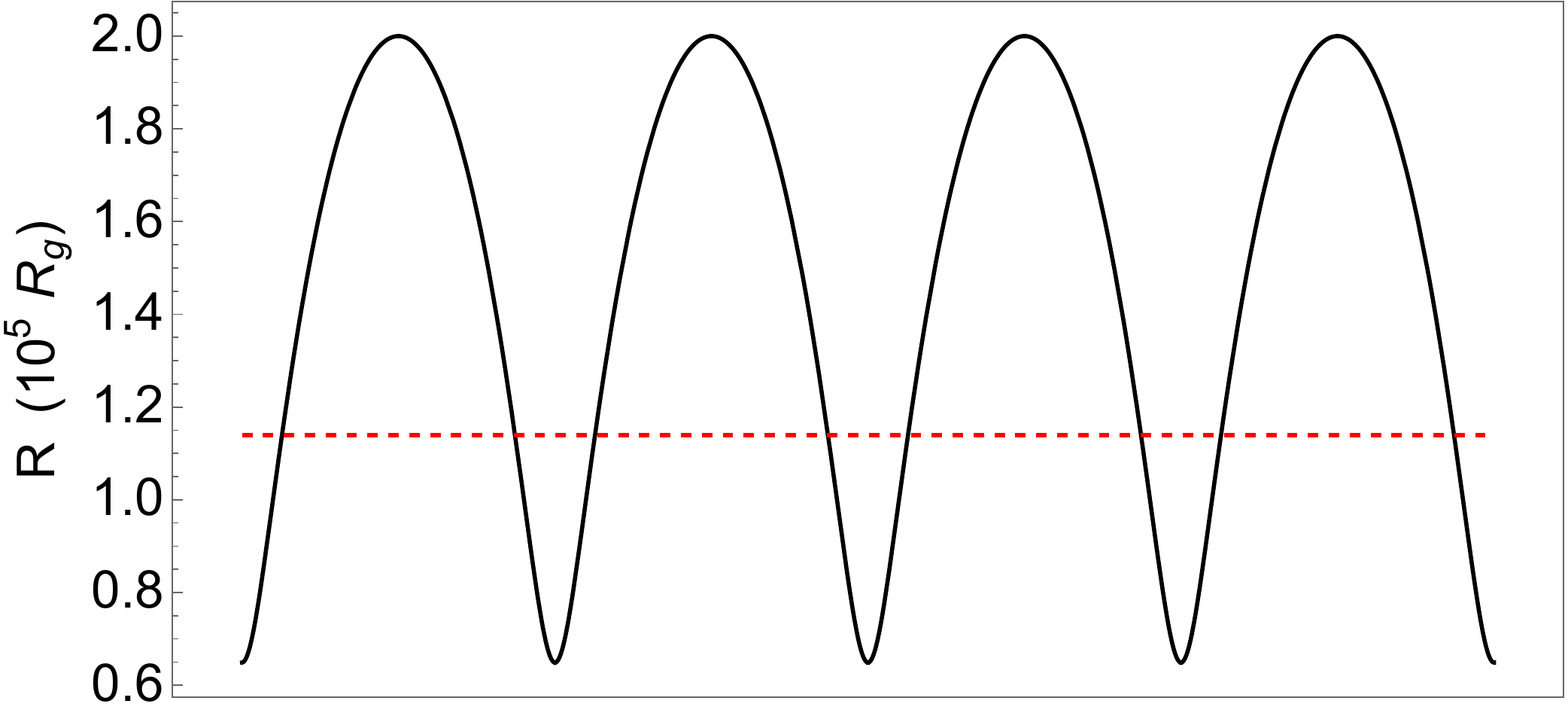}}
\vspace{0.1cm}
\hbox{\includegraphics[scale=0.4]{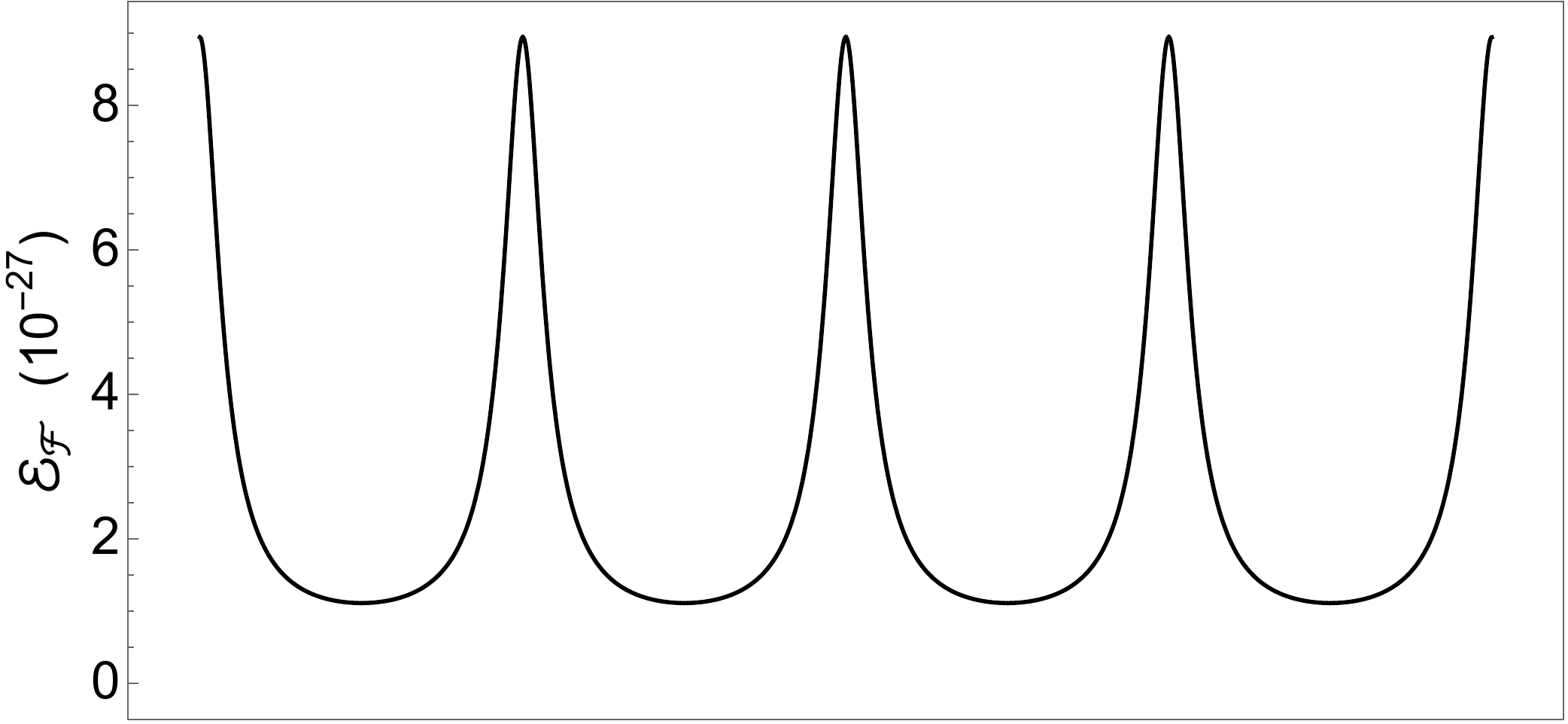}}
\vspace{0.1cm}
\hbox{\includegraphics[scale=0.4]{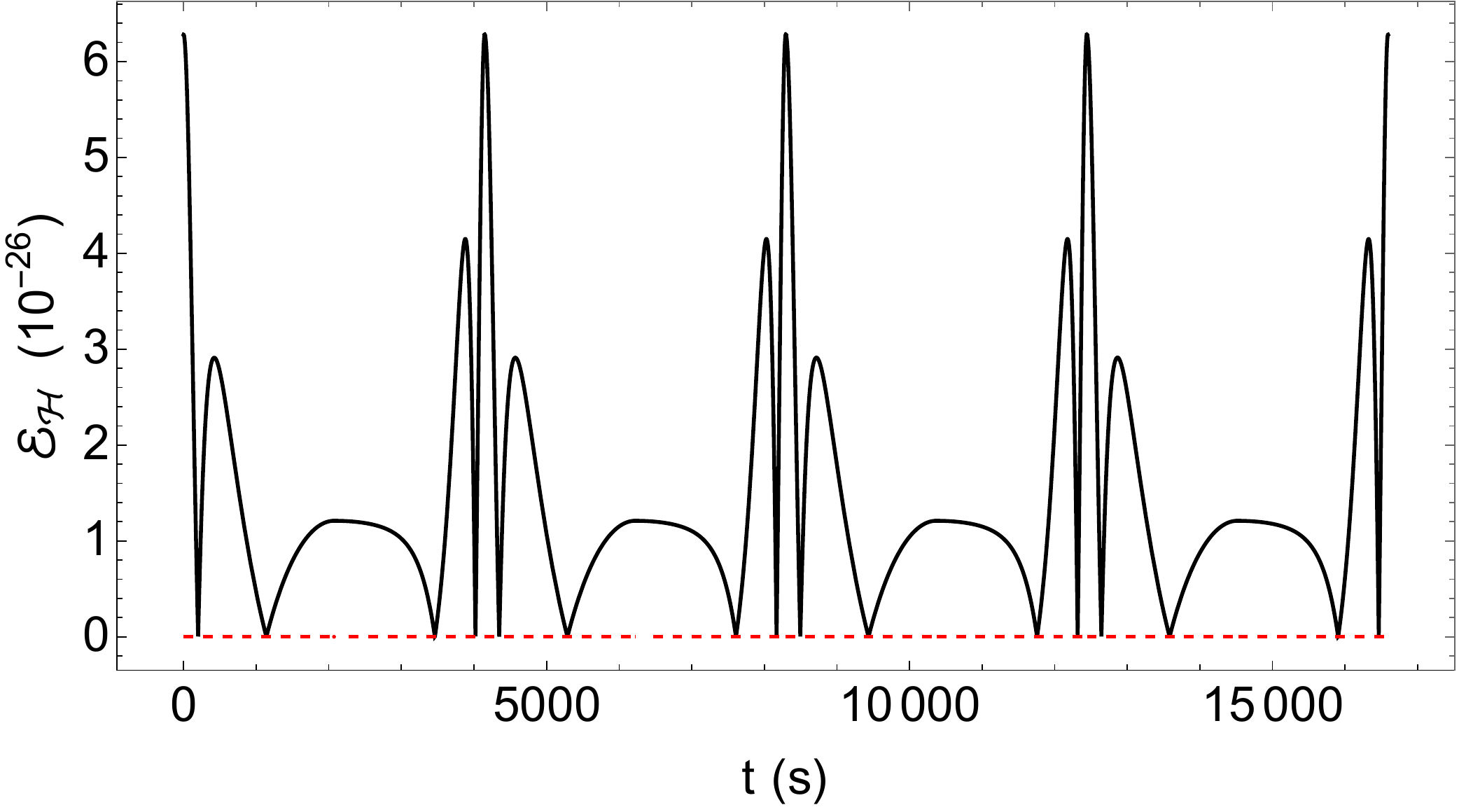}}
\caption{Plots of the functions $R(t)$, $\mathcal{E}_{\mathcal{F}}(t)$, and  $\mathcal{E}_{\mathscr{H}}(t)$. \emph{Upper panel}: time evolution of the modulus of the relative radius (cf. Eq. \eqref{eq:relative-orbit-R-varphi});  the horizontal red dashed line corresponds to its average value (see Table \ref{tab:table1}). \emph{Middle panel}:  trend of  $\mathcal{E}_{\mathcal{F}}(t)$ (see Eq. \eqref{eq:E_F}). \emph{Lower panel}: behavior of  $\mathcal{E}_{\mathscr{H}}(t)$ (cf. Eq. \eqref{eq:E_H}); the horizontal  red dashed line represents the modulus of the mean EC contribution, which amounts to $9.31\times10^{-53}$.}
\label{fig:Fig_NS}
\end{figure}
 
The order of magnitude of the spin components $s_{\rm z1}$ and $s_{\rm z2}$ (in units of $\hbar$), physically representing the number of neutrons  inside the NSs, is consistent with the values reported in the literature (which are of the order of $ 10^{57}$ neutrons) \cite{Shapiro1983}; moreover, the magnitude of the parameters $v_1^{\rm max},v_2^{\rm max},\alpha_1,\alpha_2,\gamma$ confirms that the slow-motion, wide-separation, and weak-field hypotheses are fulfilled.

\begin{figure}[h!]
    \centering
    \includegraphics[scale=0.3]{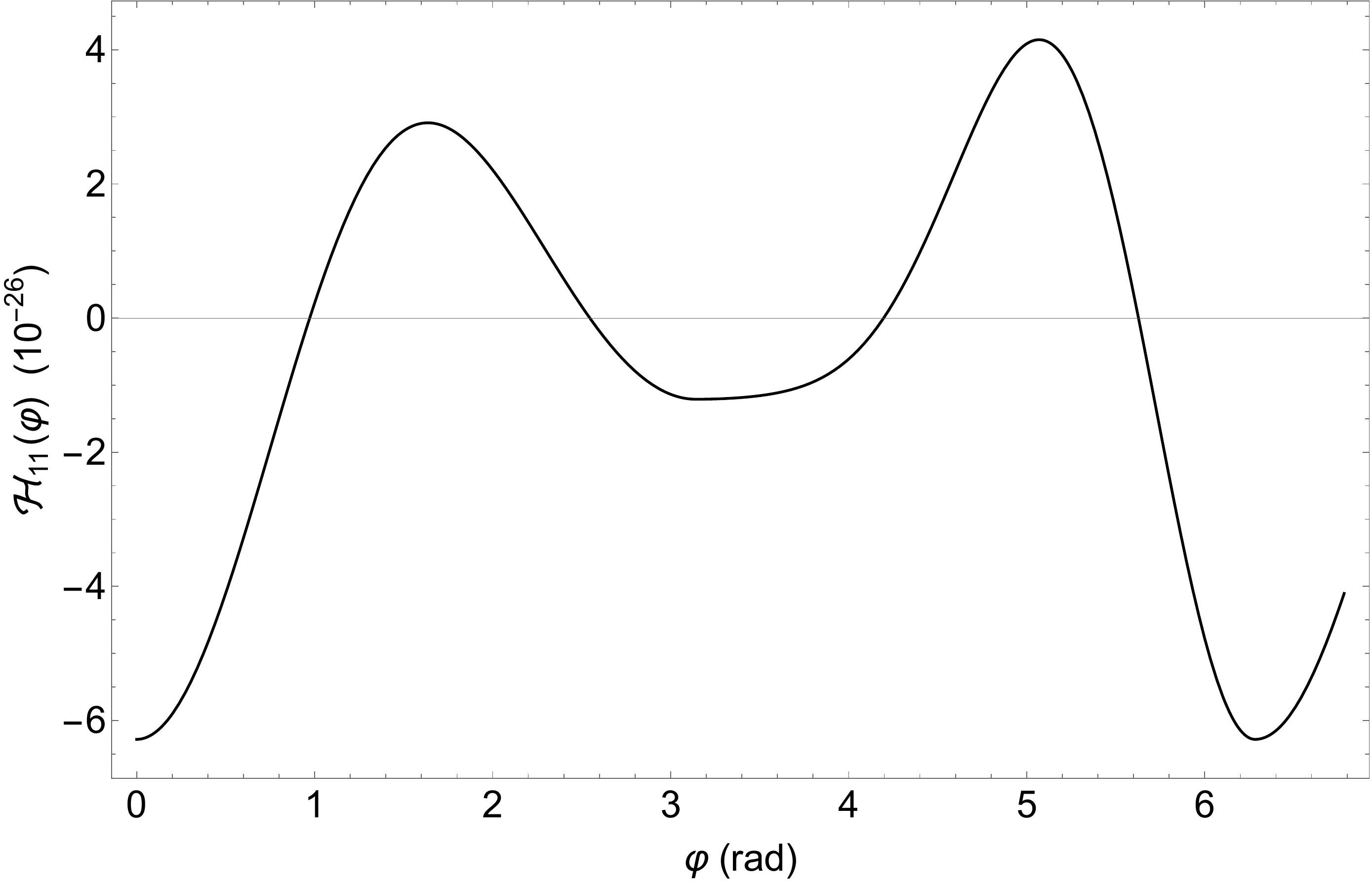}
    \caption{Gravitational waveform $\mathscr{H}_{11}$ as a function  of the azimuthal angle $\varphi$. The $\varphi$-intercepts occurr at $\varphi=0.31\pi,\ 0.81\pi,\ 1.34\pi,\ 1.79\pi$.}
    \label{fig:GWF}
\end{figure}

In order to estimate the EC contributions  to the GR flux and   waveform, we define (cf. Eqs. \eqref{eq:luminosity-plot}--\eqref{eq:H-11-EC-in-plot})
\begin{subequations}
\label{eq_E_F-and-E_H}
\begin{align}
\mathcal{E}_{\mathcal{F}}(t)&\equiv \left|\frac{\mathcal{F}_{\rm EC}(t)}{\mathcal{F}_{\rm GR}(t)}\right|,
\label{eq:E_F}\\
\mathcal{E}_{\mathscr{H}} (t)&\equiv|\mathscr{H}^{\rm GR}_{11}(t)|-|\mathscr{H}^{\rm EC}_{11}(t)|.
\label{eq:E_H}
\end{align}
\end{subequations}
The  above quantities,  along with the function $R(t)$ representing the relative distance of the NSs, are shown in Fig. \ref{fig:Fig_NS}. From the plot of $\mathcal{E}_{\mathcal{F}}$, we see that the spin effects become more significant at the closest point of approach between the objects, where the gravitational field becomes more intense. This agrees with the spirit of EC theory, whose importance is expected to increase  in the strong-gravity regime. In our example, the average contributions predicted by EC theory are smaller than GR ones by a factor of  $10^{-23}$. This difference is consistent with  the fact that  the bodies are widely separated during their dynamical evolution.  The EC  corrections can be also figured out  starting from the trend of $\mathcal{E}_{\mathscr{H}}$. As shown in Fig. \ref{fig:Fig_NS},  it  goes from  its minimum to its maximum values when the NSs get closer. Moreover, this function  vanishes as soon as $\mathscr{H}_{11}(t)=0$ (cf. Eqs. \eqref{eq:GWF-plot} and \eqref{eq:E_H}).  These points indicate when GR and EC effects become comparable and in our example we have $|\mathscr{H}_{11}^{\rm GR}| = |\mathscr{H}_{11}^{\rm EC}|\sim 10^{-52}$. The same information, expressed in terms of  $\varphi$, can be inferred from the zeroes of $\mathscr{H}_{11}(\varphi)$, which occur at $\varphi\approx \frac{1}{4}\pi,\frac{3}{4}\pi,\frac{7}{4}\pi, \frac{11}{4}\pi$ (see Fig. \ref{fig:GWF}; the functional form of $\mathscr{H}_{11}(\varphi)$ can be promptly deduced from Eqs. \eqref{eq:GWF-plot}--\eqref{eq:H-11-EC-in-plot}).

\section{Conclusions}
\label{sec:end}

This work configures as a natural continuation of the research program started out in Ref. \cite{Paper1}, where we have solved the GW generation problem in EC model at 1PN level by resorting to the Blanchet-Damour formalism. This general treatment finds an  explicit application here, where the matter source is described by the Weyssenhoff fluid.

\begin{figure*}[ht!]
    \centering
    \includegraphics[scale=0.53]{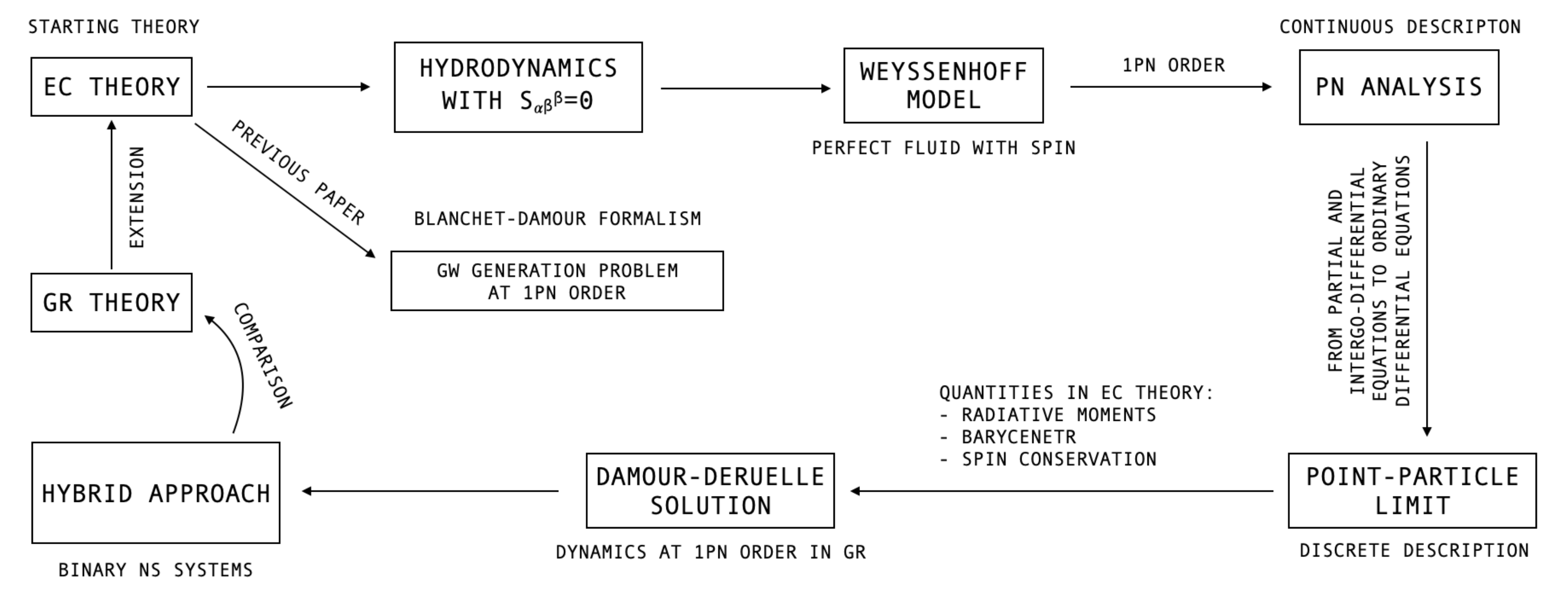}
    \caption{Flowchart of the paper. \qm{Previous paper} refers to Ref. \cite{Paper1}.}
    \label{fig:scheme}
\end{figure*}

The  structure of the paper is sketched in Fig. \ref{fig:scheme}. In Sec. \ref{Sec:1PN-generation-GWs-EC}, we have summarized the key steps of the previous article. In this framework, the  spinning PN source is supposed to be a generic hydrodynamical   fluid system. For this reason, in Sec. \ref{sec:rel_hyd}, we have  introduced the fundamental pillars of the EC hydrodynamics verifying the simplifying hypothesis that the torsion tensor has a vanishing trace, i.e., $S_{\alpha\beta}{}^\beta=0$. Subsequently,  we have  modeled  the spin effects inside  matter by employing the Weyssenhoff model of a semiclassical ideal spinning fluid supplemented by the Frenkel condition (see Sec. \ref{sec:Weyssenhof}). The  study of the Weyssenhoff fluid  within the PN approximation scheme, representing a  fundamental  tool of the Blanchet-Damour formalism, is contained in Sec.  \ref{Sec:PN-Hydrodynamics}. Both at 0PN and at 1PN level, the dynamics is ruled by a system of partial and integro-differential equations, whose resolution is extremely demanding. A less involved pattern can be obtained  if we  employ the point-particle procedure, which allows to characterize the fluid dynamics in terms of ordinary differential equations 
by  going from a continuous picture to a discrete description of the system (see Sec. \ref{sec:Point-particle-limit}). We have then derived, within EC theory and for the particular case of binary systems,  the 1PN formula \eqref{eq:center-of-mass-2-bodies} of  the  center of mass position  and the  general expressions \eqref{I-ij-rad-pp-limit-expression-2}--\eqref{J-ijk-rad-pp-limit-expression-2}
of the radiative multipole moments. Starting from these results and the conservation law \eqref{eq:spin-conserved-expr} of the spin vector, we have  resorted to the  Damour-Deruelle solution in GR (which has been briefly discussed in    Sec. \ref{sec:motion_1PN_GR}) to  set up a hybrid approach for dealing with binaries of spinning PN NSs, where we have provided some numerical estimates of the EC contributions to the flux and the waveform (see Secs. \ref{sec:flux-gwf} and \ref{sec:num_sim}).

This paper contains some new theoretical results,  which can be summarized as follows:
\begin{enumerate}
\item[(1)] development of a general pattern (subject to the hypothesis $S_{\alpha\beta}{}^\beta=0$) for the hydrodynamics in EC theory, where the spin effects are modeled through the tensors  $\Phi^{\alpha\beta}$ and  $\tau_{\mu\nu}{}^\lambda$ (cf. Eq. \eqref{system of general matter equations});
\item[(2)] PN investigation of the Weyssenhoff fluid, which predicts at 0PN level that: $(a)$ the translational motion  matches the Newtonian Euler equation (see Eq. \eqref{eq:0PN-eq2}); $(b)$ the rotational dynamics  reduces to a homogeneous continuity equation (see Eq. \eqref{eq:0PN-eq3}); $(c)$ the expressions  of the luminosity and the gravitational waveform  reproduce formally the corresponding  GR quadrupole formulas (cf. Eqs. \eqref{eq:power-radiated-1PN_equivalent-expr} and  \eqref{eq:gravitational_wave_amplitude-equival-express});
\item[(3)] derivation of the  radiative multipole moments for the Weyssenhoff fluid  both in the fine-grained and the coarse-grained (for an $N$-body  and a binary system) descriptions, see Eqs.  \eqref{eq:I-ij-rad-expression-1}--\eqref{eq:J-ijk-rad-expression-1}, \eqref{eq:I-ij-rad-point-particle-limit}--\eqref{eq:J-ijk-rad-point-particle-limit}, and \eqref{I-ij-rad-pp-limit-expression-2}--\eqref{J-ijk-rad-pp-limit-expression-2};
\item[(4)] computation,  for the first time in the literature, of the numerical value of the NS spin as conceived in the  Weyssenhoff semiclassical model  (cf. Eq. \eqref{eq:szi-components});
\item[(5)] hybrid scheme for providing a first estimate of the EC contributions to the GWs emitted by a binary NS system.
\end{enumerate}

The calculation of the parameter  $n$ appearing in the  spin formula \eqref{eq:szi-components} has led to established results for the number of neutrons inside a NS. Moreover, our investigation has revealed that EC contributions become more important when the gravitational field strength grows (see Fig. \ref{fig:Fig_NS}). Therefore, we could in principle  extend our hybrid scheme to binary BH systems and evaluate the corrections to the radiated power and the asymptotic amplitude  foretold by EC model. If we suppose that, similarly to Eq.  \eqref{eq:szi-components},  the BH spin can  be written as
\begin{align}
\label{eq:szi-components_BH}
s=  n\hbar\frac{4\pi}{3}\left(\frac{2Gm}{c^2}\right)^3,
\end{align}
($m$ being the BH mass) then we obtain that the EC effects are, for known astrophysical masses (i.e., $6 M_{\odot}\lesssim M \lesssim 10^{10} M_{\odot}$,  with $M$  the total mass of the system), between  23 and 13 orders of magnitude lower than GR ones. These differences  are justified by the fact that our approach is restricted to compact binaries in their early inspiralling stage. Despite that, our  analysis  permits to infer  that EC corrections can fulfill a relevant role in the later evolution phases. This topic might be useful for testing quantum phenomena in the strong-gravity regime, and it is worth examining it in a separate paper. Furthermore, GW phenomena could  lead to interesting implications in the context of generalized EC theories. In particular,  Ref. \cite{Stornaiolo2012} puts forth a model where the gravitational Lagrangian has an additional quadratic-torsion term whose coupling constant differs from the Newtonian gravitational constant $G$. This new  contribution changes the interaction strength  between  torsion and matter fields and hence novel results can be found also in the GW framework. 

In the context of the GR pole-dipole approximation, the lowest-order \qm{classical spin} contributions to the mass-type and current-type radiative  moments emerge at 1.5PN and 0.5PN level, respectively \cite{Kidder1995,Blanchet2014}. Therefore, in EC theory we recover \emph{formally} the same expressions as in GR, see Eqs. \eqref{I-ij-rad-pp-limit-expression-2}--\eqref{J-ijk-rad-pp-limit-expression-2}. This is a consequence of  the Frenkel condition \eqref{eq:FC_leading_next_to_leading_2} and the ensuing PN expansions of the tensors $T^{\mu \nu}$ and $\mathcal{S}^{\mu \nu}$ (see Eqs. \eqref{eq:PN-expansion-T-alpha-beta} and \eqref{eq:PN-expansion-S-alpha-beta}).  If we define the Kerr angular momentum as $\mathcal{J}=\tfrac{Gm^2}{c}a$ with $a\in(0,1)$, we can provide an estimate of the effects introduced by EC theory on a single body by evaluating the ratio $s/\mathcal{J}$ by means of Eqs. \eqref{eq:szi-components} and  \eqref{eq:szi-components_BH}. If we choose $a=0.5$, in the   NS case we have  $s/\mathcal{J}\sim 10^{-11} $ for $m\in[1.1,2.2]\ M_\odot$, while for a BH $s/\mathcal{J}\sim(10^{-12}-10^{-2})$ with $m\in[3,10^{11}]\ M_\odot$.

In conclusion, this paper  prepares the ground for a systematic study of spinning PN binaries in EC theory. This  entails a comprehensive investigation of  the 1PN dynamics of binary systems in EC theory, which deserves consideration in a separate paper.

\section*{Acknowledgements}
The authors are grateful to C. Stornaiolo and L. Stella for extensive discussions. E. B. is grateful to F. W. Hehl and Y. N. Obukhov for valuable correspondence. V. D. F. thanks D. Perrodin and A. Ridolfi for useful discussions.  The authors thank  Gruppo Nazionale di Fisica Matematica of Istituto Nazionale di Alta Matematica for partial support, and  the Silesian University in Opava and the International Space Science Institute in Bern for hospitality and partial support. V.D.F. thanks the University of Vienna for hospitality and partial support. V. D. F. acknowledges the support of INFN {\it sez. di Napoli}, {\it iniziative specifiche} TEONGRAV. E.B. acknowledges the support of  the Austrian Science Fund (FWF) grant P32086. E. B. dedicates this paper to his beloved grandmother Emilia.

\appendix
\section{Post-Newtonian expansion of the Riemann tensor}
\label{sec:Riemann}

We report selected PN expansions of the Riemann tensor components (see Eq. \eqref{eq:EC-Riemann-tensor}). These are given by
\begin{align}
{}^{(3)}R^k_{\ l0i}&= \partial_i\left(\frac{\chi}{2} c\,{}^{(1)}s_{kl}+\dfrac{4}{c^3}\partial_{[l}V_{k]}\right)\notag\\
&+\dfrac{2}{c^3}\delta_{i[k}\partial_{l]}\partial_t V,
\\
{}^{(2)}R^k_{\ lij}&= \dfrac{2}{c^2}\left(\delta_{k[j}\partial_{i]}\partial_l+\delta_{l[i}\partial_{j]}\partial_k\right)V,
\\
{}^{(2)}R^k_{\ 00i}&=\dfrac{1}{c^2} \partial_i \partial_{k}V,
\label{eq:PN_Riemann_3_term}
\\
{}^{(4)}R^k_{\ 00i}&=-\chi\left[
\partial_i\left({}^{(1)}s_{kp}v^p\right)+\frac{1}{2}\partial_t({}^{(1)}s_{ki})\right]\notag\\
&+\frac{1}{c^4}\delta_{ki}\left[\partial_lV\partial_lV+\partial_t\partial_tV\right]\notag\\
&-\frac{1}{c^4}\left[4V\partial_k\partial_iV+3\partial_kV\partial_iV\right.\notag\\
&\left.+2\partial_t(\partial_kV_i+\partial_iV_k)\right],\\
{}^{(3)}R^k_{\ 0ij}&= \chi\, c\partial_{[j} {}^{(1)}s_{k|i]}\notag\\
&+\frac{2}{c^3}\left(2\partial_k \partial_{[i} V_{j]}+\delta_{k[j}\partial_{i]}\partial_tV\right),
\end{align}
where ${}^{(n)}R^{\mu}_{\ \nu\alpha\beta}\sim\left(\dfrac{\bar{v}}{c} \right)^n\dfrac{1}{\bar{d}^2}$ and we have used the Frenkel condition \eqref{eq:FC_leading_next_to_leading_2}. The above equations give contributions in Eq. \eqref{eq:1PN-Euler-equation-explicit}.

\end{document}